\documentclass[12pt]{myiopart}

\usepackage{axodraw}
\usepackage{graphicx}
\usepackage{rotating}
\usepackage{color}

\begin{document}




\title{Searches for Leptoquarks with the ATLAS Detector}

\author{Andr\'e Sopczak on behalf of the ATLAS Collaboration}

\address{Institute of Experimental and Applied Physics, Czech Technical University in Prague}
\ead{andre.sopczak@cern.ch}

\begin{center}
{\bf\small ABSTRACT}
\end{center}
{\large 
Leptoquarks (LQ) are predicted by many new physics theories to describe the similarities between the lepton and quark sectors of the Standard Model and offer an attractive potential explanation for the lepton flavour anomalies observed at LHCb and flavour factories. The ATLAS experiment has a broad program of direct searches for Leptoquarks, coupling to the first-, second- or third-generation particles. The most recent 13\,TeV results on the searches for Leptoquarks and contact interactions with the ATLAS detector are reviewed, covering flavour-diagonal and cross-generational final states.
}

\vspace*{8cm}
\begin{center}
{\em Contribution to the 
Phenomenology 2021 Symposium, \\
Pittsburgh, online, 24-26 May 2021}
\end{center}

\maketitle
\setcounter{page}{1}

\setlength{\textheight}{250mm}

\newcommand{\LQu}{\mbox{$\mathrm{LQ^u}$}}
\newcommand{\LQd}{\mbox{$\mathrm{LQ^d}$}}
\newcommand{\LQuthree}{\mbox{$\mathrm{LQ^u_3}$}}
\newcommand{\LQdthree}{\mbox{$\mathrm{LQ^d_3}$}}
\newcommand{\LQumix}{\mbox{$\mathrm{LQ^u_{mix}}$}}
\newcommand{\LQdmix}{\mbox{$\mathrm{LQ^d_{mix}}$}}

\newcommand{\bb}{\mbox{$\mathrm{b\bar b}$}}
\newcommand{\toptop}{\mbox{$\mathrm{t\bar t}$}}
\newcommand{\nn}{\nu\nu}
\newcommand{\tautau}{\tau^+\tau^-}
\newcommand{\ee}{\mbox{$\mathrm{e}^{+}\mathrm{e}^{-}$}}

\newcommand{\Zo} {{\mathrm {Z}}}
\newcommand{\db}    {{d_{\rm B}}}
\newcommand{\dgz}  {{\Delta g_1^{\Zo}}}
\newcommand{\dkg}   {{\Delta \kappa_\gamma}}

\newcommand{\pb}   {\mbox{$\rm pb^{-1}$}}
\newcommand{\fb}   {\mbox{$\rm fb^{-1}$}}

\clearpage
\section{Introduction}

\begin{figure}[htb]
\vspace*{-0.2cm}
\begin{center}
\includegraphics[width=\textwidth,height=5cm]{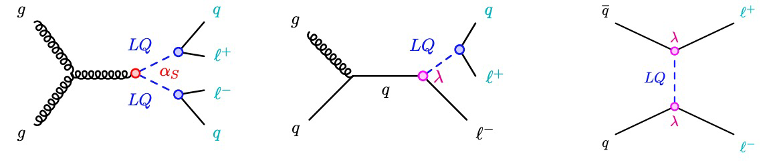}
\end{center}
\vspace*{-0.8cm}
\caption{
\newcommand{\Cc}{\ensuremath{\kappa}}
Feynman diagrams for Leptoquark production.
Left: 
pair-production.
Centre:
single-production.
Right:
off-shell production.
}
\label{fig:production}
\vspace*{-0.1cm}
\end{figure}

Leptoquarks are receiving increasing interest with the observation of 
flavour anomalies~\cite{lhcbcollaboration2021test}.
The Leptoquarks are colour triplet bosons with fractional charge,
and they decay flavour-diagonal and also possibly cross-generations.
Their Yukawa interaction can be described with a coupling $\lambda$. 
There are three main production modes, pair-production, single-production and	
off-shell production, as illustrated in Figure~\ref{fig:production}.
The pair-production has a large cross-section, 
while the single-production is 
sensitive for large masses and has a 
cross-section proportional to $\lambda^2$,
and the off-shell production is sensitive to
even larger masses with a 
cross-section proportional to $\lambda^4$.

Table~\ref{tab:decay} lists the definition of 
the branching fraction $B$ for the LQ decays.

\begin{table}[htb]
\vspace*{-0.2cm}
\caption{
Branching fraction $B$ for the LQ decays.
\label{tab:decay} }
\begin{center} 
\vspace*{-0.1cm}
\begin{tabular}{ccc} \hline
LQ decay        & $B=1 $ & $ B=0$ \\ \hline
\LQu &$\mathrm{b\tau}$ &$\mathrm{t\nu}$   \\ 
\LQd &$\mathrm{t\tau}$ &$\mathrm{b\nu}$   \\\hline
\end{tabular} 
\end{center}
\end{table}
Motivated by observed B-anomalies,
a dedicated search for bs$\ell\ell$
in data recorded by the ATLAS detector~\cite{Aad:1129811} is given in Section~\ref{sec:bsll}.
Third Generation LQs are 
of highest interest and they are 
addressed in Section~\ref{sec:thrid}
for a signature of b-jets
(\bb), missing transverse energy (MET) and a $\tau$-lepton.
Searches for pair-production of third-generation down-type Leptoquarks in \bb+MET events are discussed in Section~\ref{sec:third_down}.
Searches for \toptop+MET
with all-hadronic final states are addressed in Section~\ref{sec:ttMEThad}.
A summary of ATLAS results is 
given in Section~\ref{sec:summary} for
up-type third-generation (\LQuthree),
down-type third-generation (\LQdthree),
up-type mixed-generation (\LQumix), and 
down-type mixed-generation  (\LQdmix) models, together 
with a review of previous ATLAS Leptoquark results.

\clearpage
\section{Search of bs$\mathbf{\ell\ell}$}
\label{sec:bsll}
\begin{figure}[htb]
\vspace*{-0.4cm}
\begin{center}
\includegraphics[width=\textwidth,height=4cm]{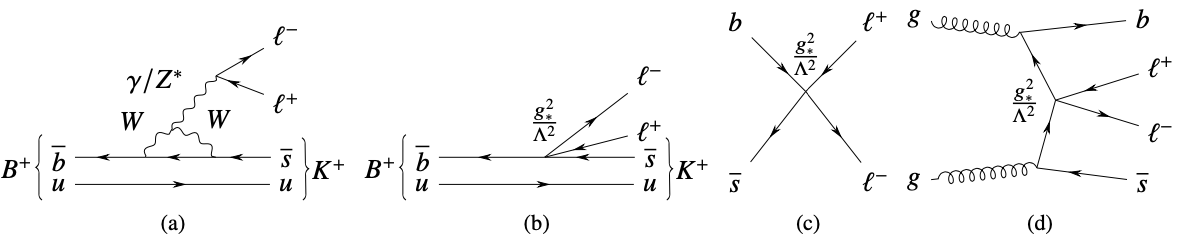}
\end{center}
\vspace*{-0.8cm}
\caption{
Representative Feynman diagrams for the decay of a B$^+$ meson to a K$^+$ meson in association with two leptons (a) in the SM and (b) in the EFT approach, and for production of two leptons via a bs$\ell\ell$ contact interaction in pp collisions (c) without and (d) with a b-jet in the final state.~\cite{atlascollaboration2021search}
}
\label{fig:bsll}
\vspace*{-0.1cm}
\end{figure}

The ATLAS collaboration performed a dedicated search for
the bs$\ell\ell$ process~\cite{atlascollaboration2021search}, motivated
by the LHCb measurement~\cite{lhcbcollaboration2021test},
as illustrated in Figure~\ref{fig:bsll}~\cite{atlascollaboration2021search}.
The model is characterized by the energy scale and coupling, $\Lambda$ and $g$.

\begin{figure}[htb]
\vspace*{-0.2cm}
\begin{center}
\includegraphics[width=0.49\textwidth,height=5cm]{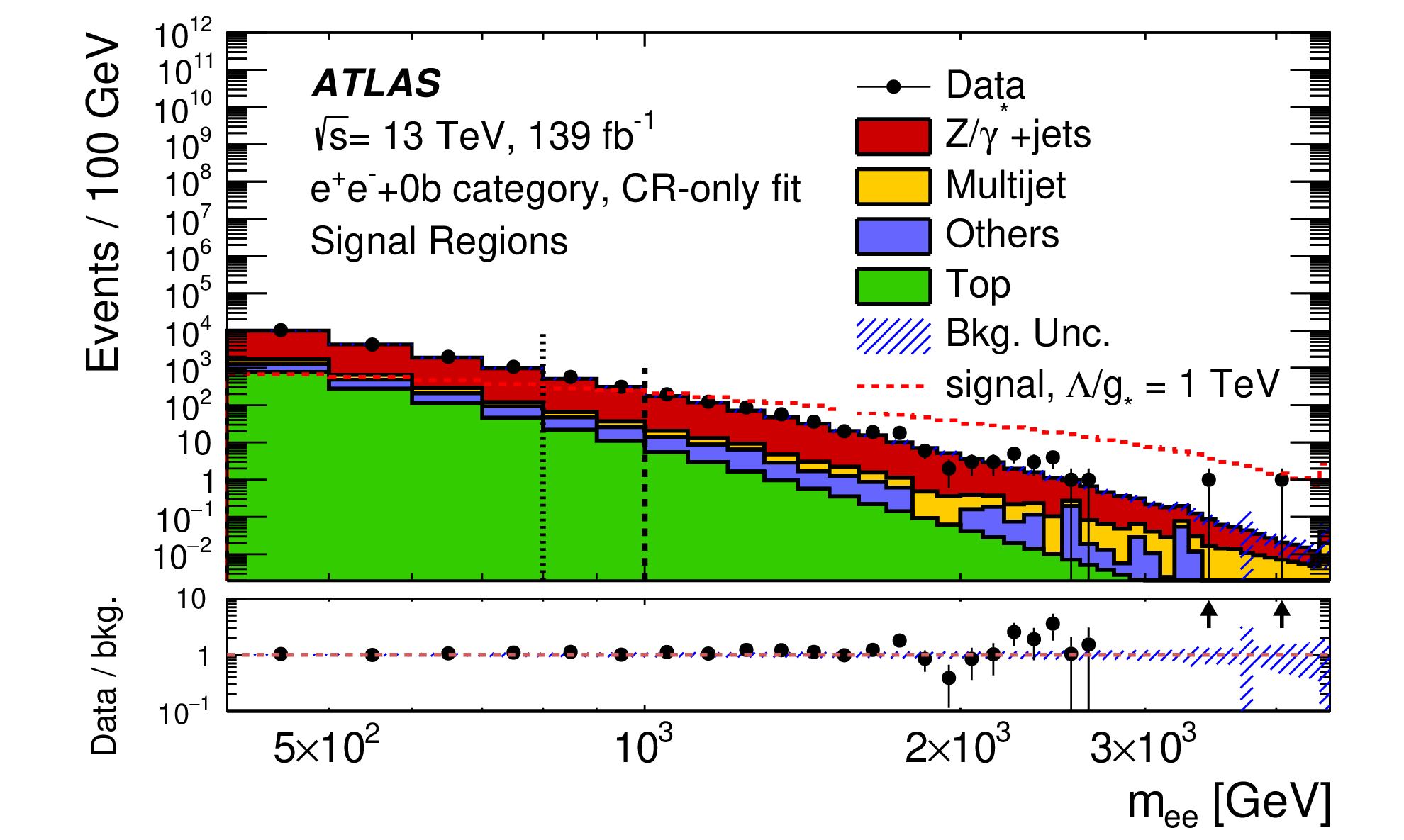}
\includegraphics[width=0.49\textwidth,height=5cm]{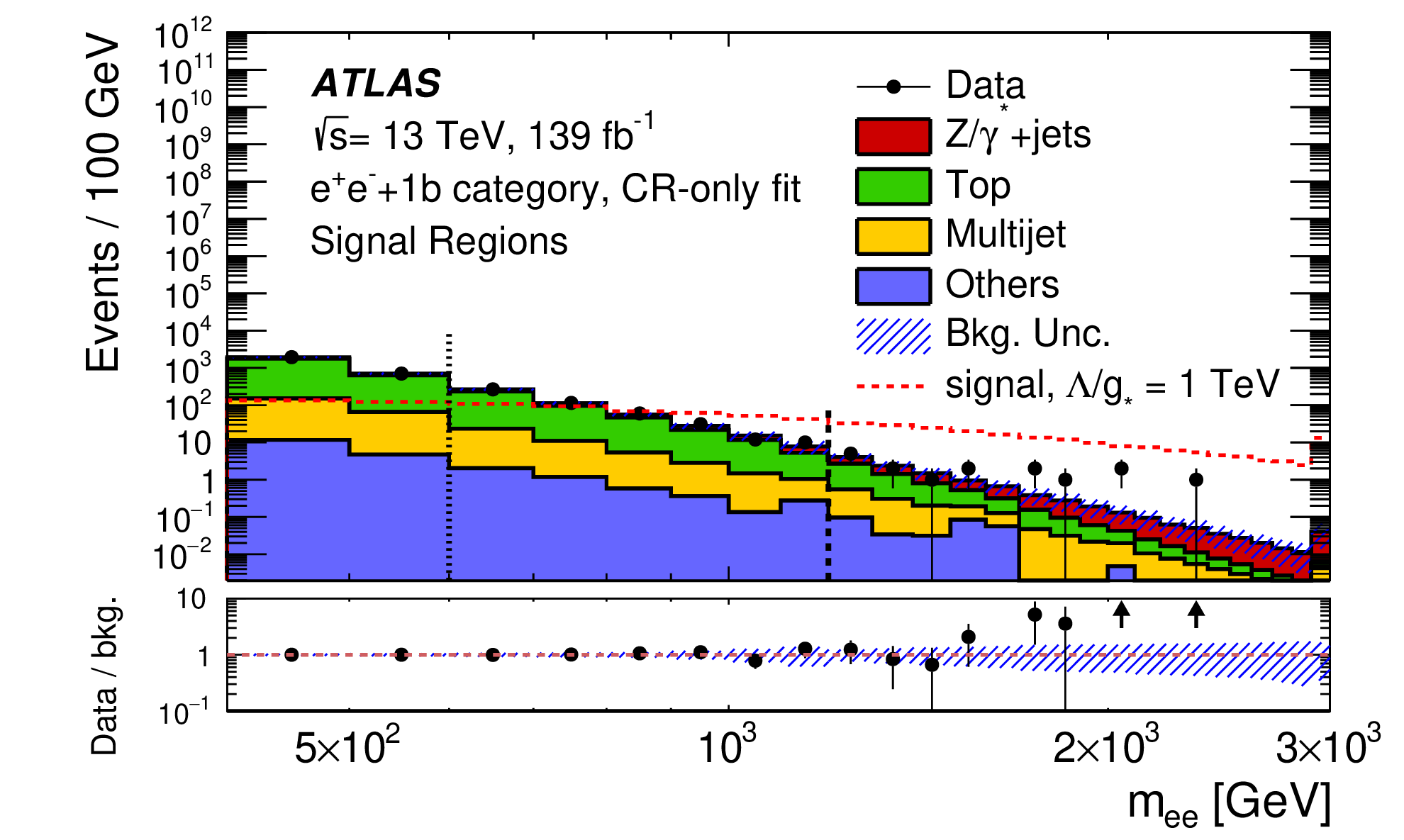}
\includegraphics[width=0.49\textwidth,height=5cm]{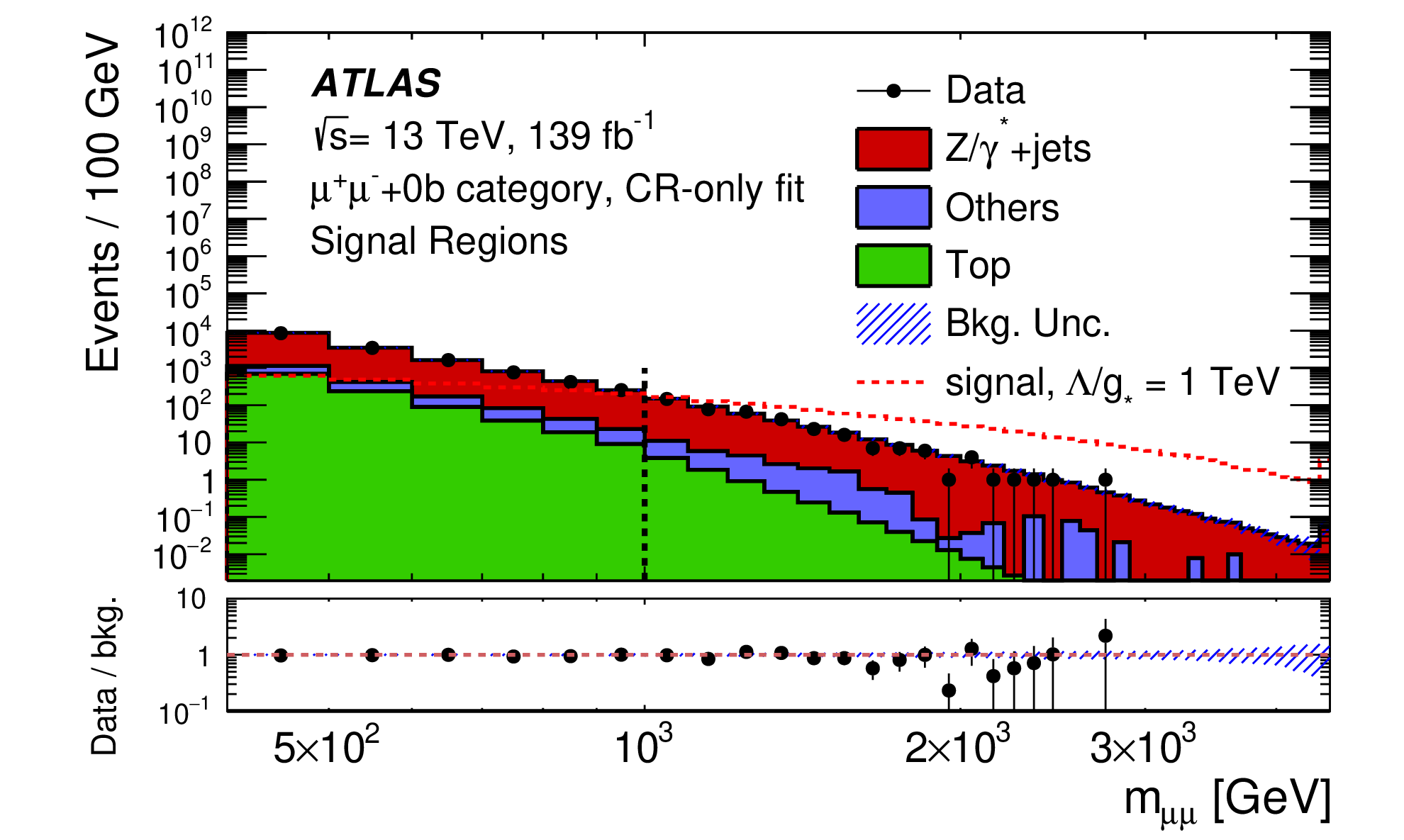}
\includegraphics[width=0.49\textwidth,height=5cm]{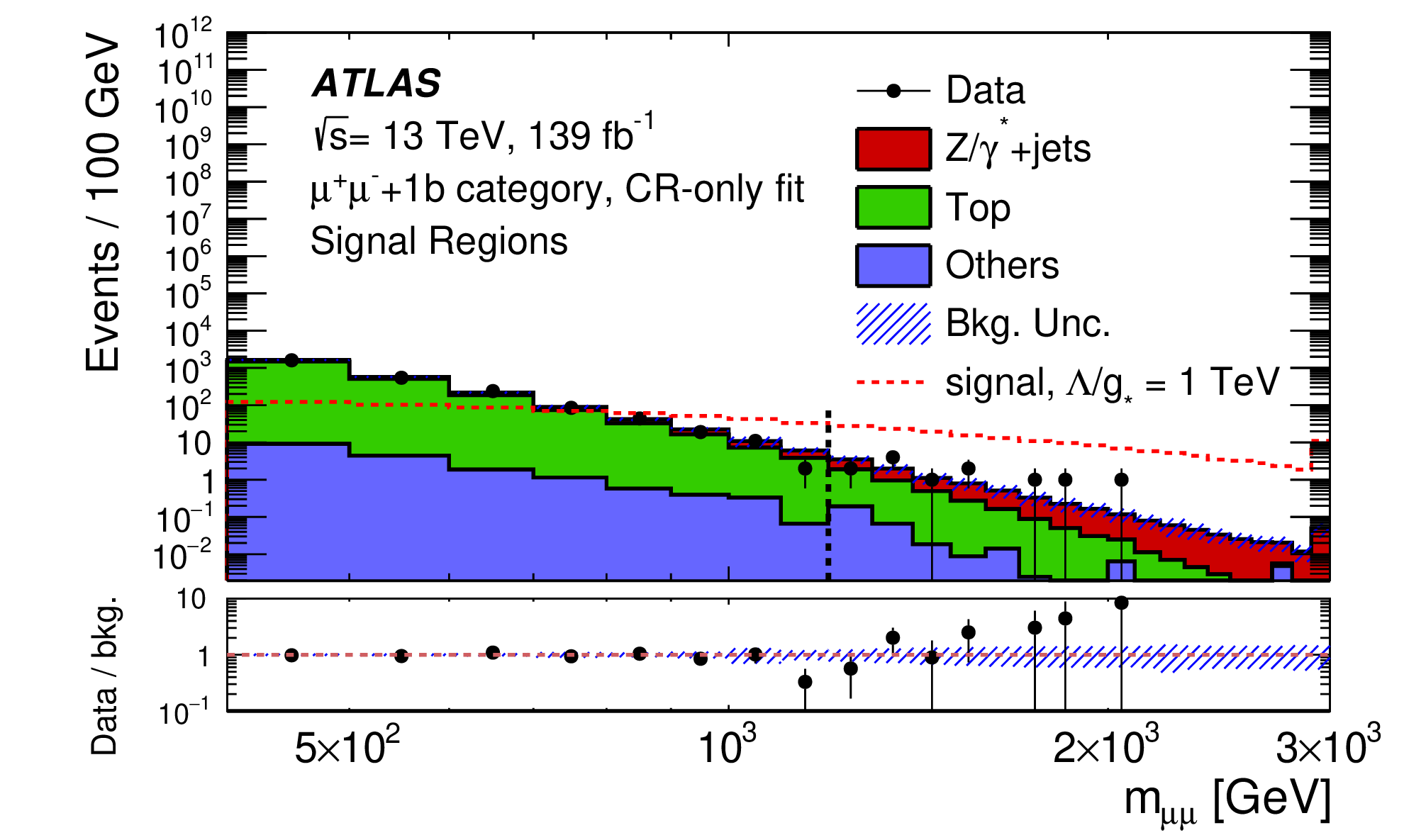}
\end{center}
\vspace*{-0.8cm}
\caption{
Data overlaid on SM background post-fit $m_{\ell\ell}$ distributions in the SRs of the (upper left) electron b-veto, (upper right) electron b-tag, (lower left) muon b-veto and (lower right) muon b-tag categories. ``Other'' refers to diboson and W+jets events. MC statistical uncertainties and systematic uncertainties are considered (hatched band). The pre-fit signal distribution is presented as well for a hypothesis of $\Lambda/g = 1$ TeV. The bottom panels show the ratio of the data to the background prediction, while the arrows correspond to bins where the ratio is beyond the limits of the figure. The last bin is an overflow bin, which contains the yields in the bins beyond it. The dashed and dotted lines mark the transition point where the extrapolation is used in the analysis for the Top and Multijet backgrounds, respectively.~\cite{atlascollaboration2021search}
}
\label{fig:bsllSM}
\end{figure}

The good agreement between data and Standard Model (SM) background is 
shown in Figure~\ref{fig:bsllSM}~\cite{atlascollaboration2021search}
for the selection with
ee0b, ee1b, $\mu\mu$0b, $\mu\mu$1b. 
There is no indication in the data for a Leptoquark signal.

\begin{figure}[htb]
\vspace*{-0.3cm}
\begin{center}
\includegraphics[width=0.49\textwidth,height=5cm]{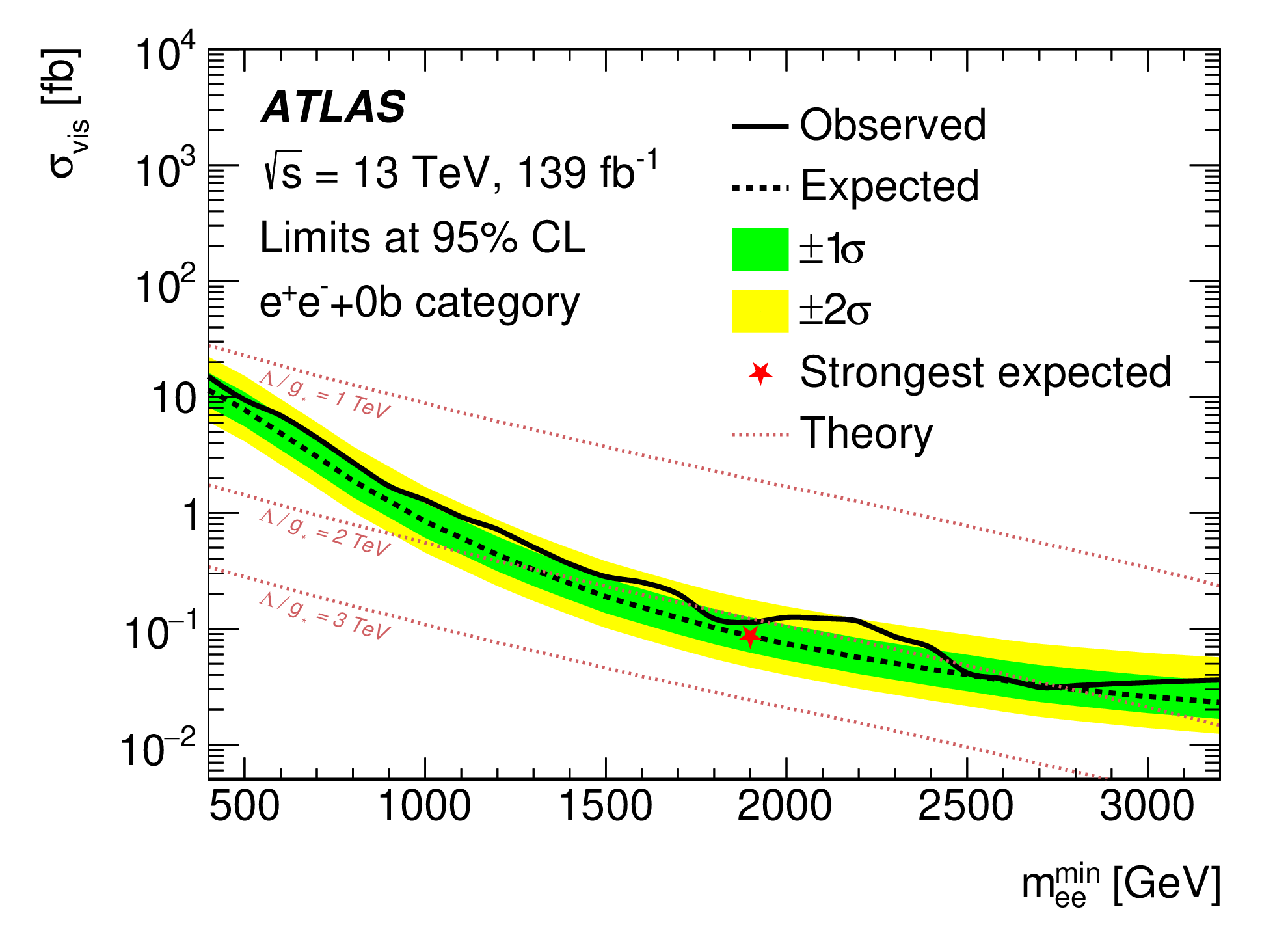}
\includegraphics[width=0.49\textwidth,height=5cm]{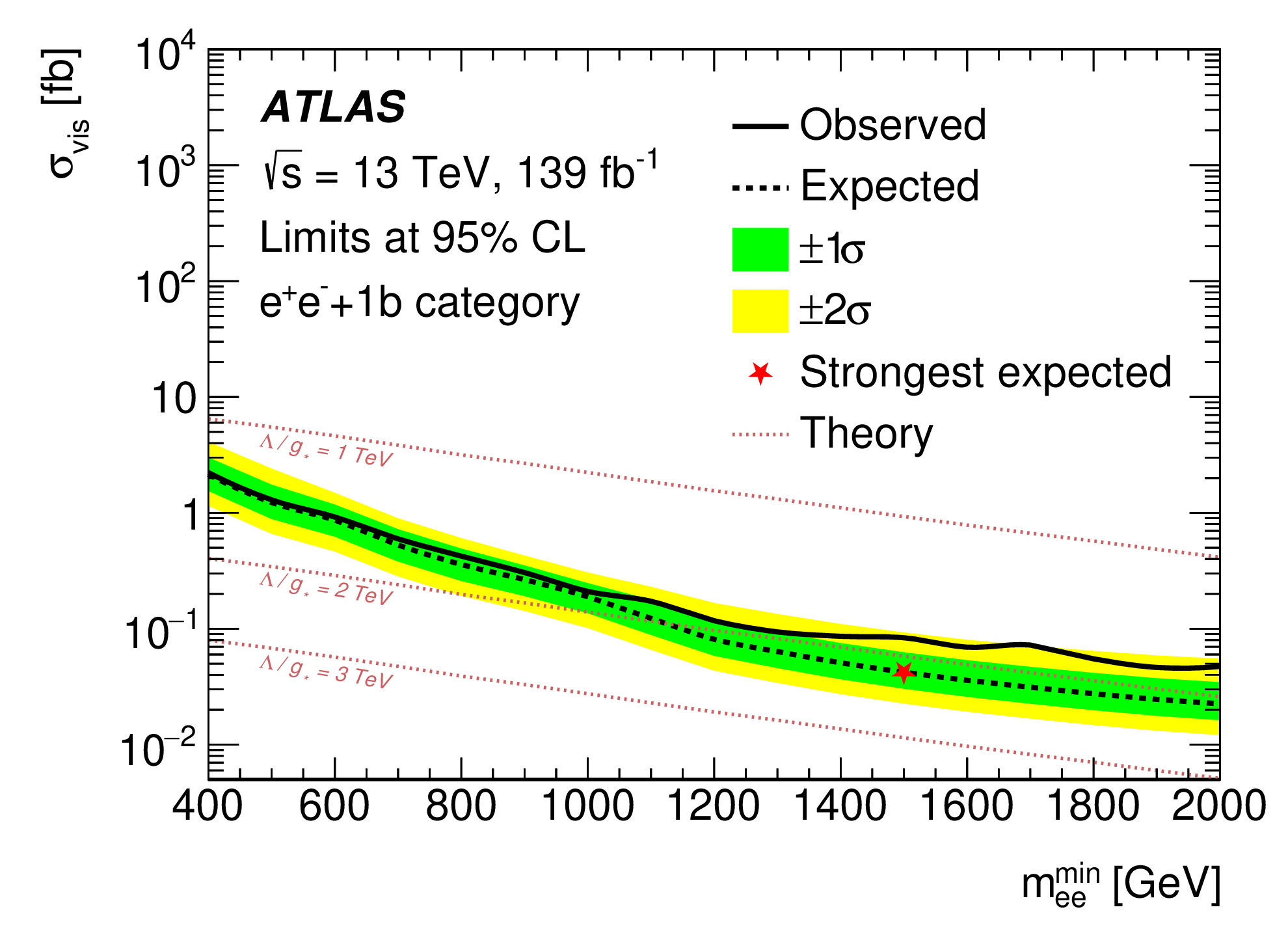}
\includegraphics[width=0.49\textwidth,height=5cm]{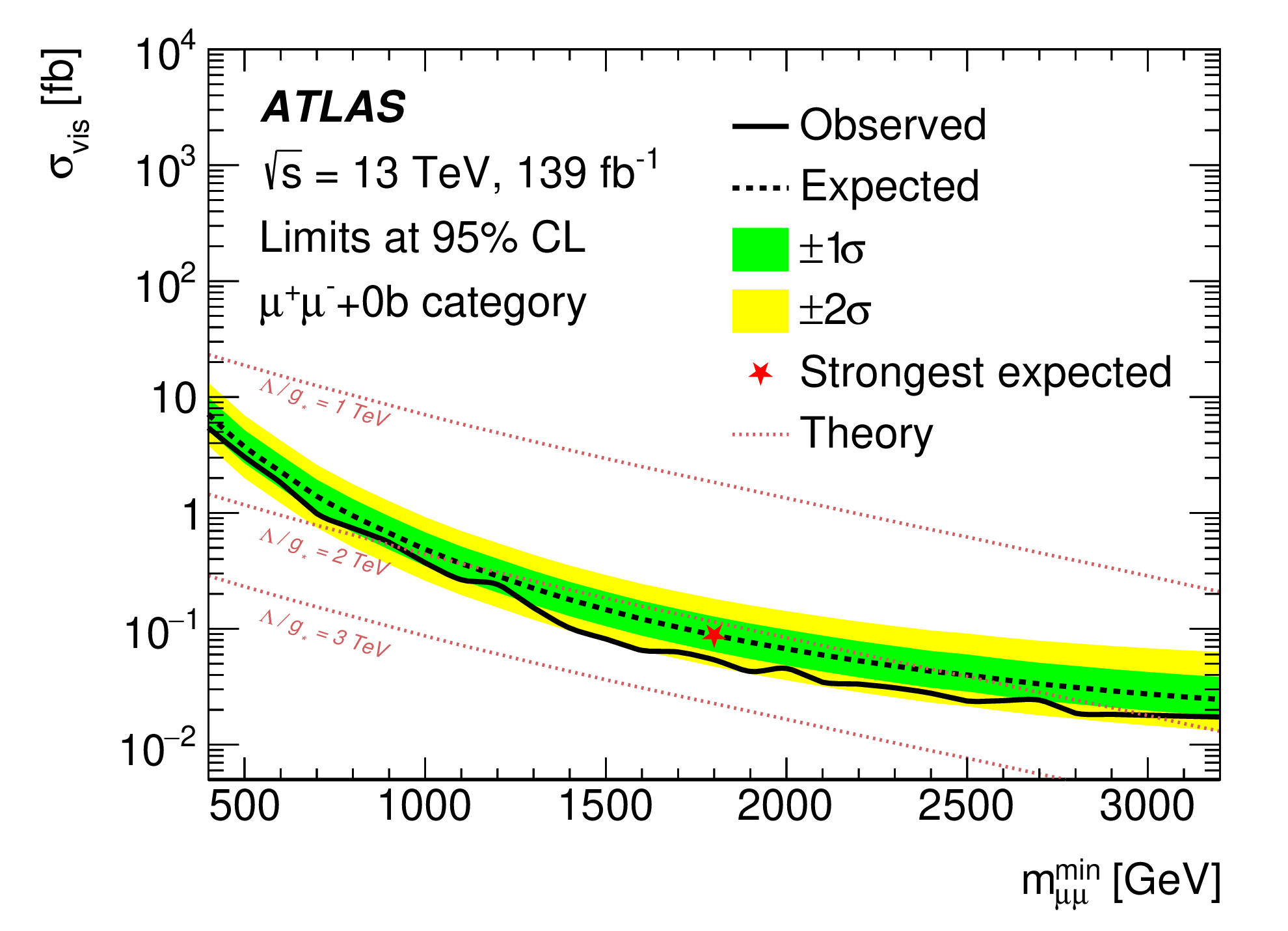}
\includegraphics[width=0.49\textwidth,height=5cm]{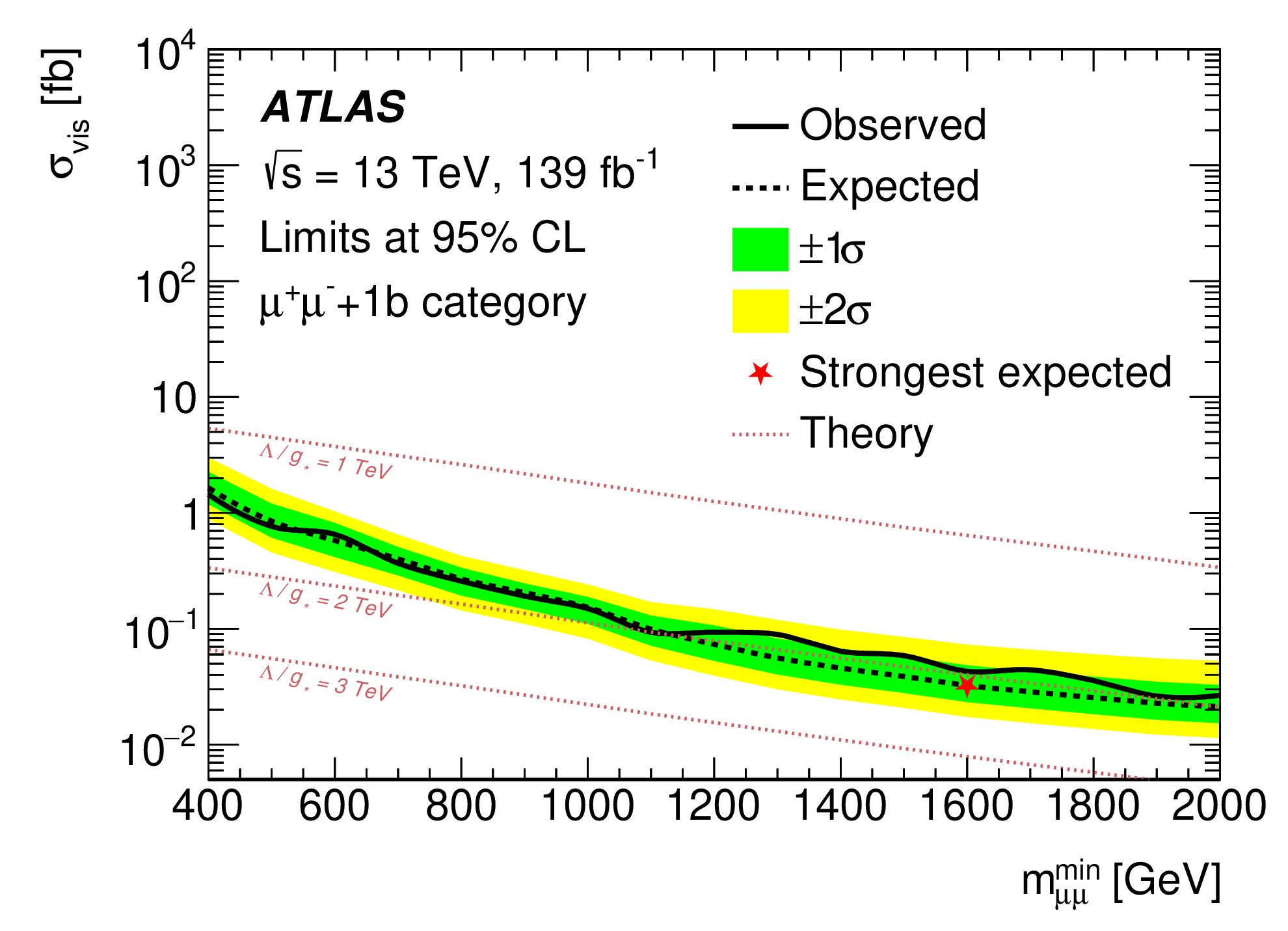}
\end{center}
\vspace*{-0.8cm}
\caption{
Model-independent observed (solid line) and expected (dashed line) upper limit on the visible cross-section 
($\sigma_\mathrm{vis} = 
\sigma\cdot\epsilon\cdot A$) for the (upper left) electron b-veto, (upper right) electron b-tag, (lower left) muon b-veto and (lower right) muon b-tag categories. The uncertainty bands around the expected limit represent the 68\% and 95\% confidence intervals. The theory lines (dotted lines) correspond to particular $\lambda/g$ values of the signal model, and the red marker presents the strongest expected lower limit on 
$\lambda/g$.~\cite{atlascollaboration2021search}
}
\label{fig:bsllLimits}
\end{figure}

\clearpage
The resulting cross-section limits at 95\%\,Confidence Level (CL) are given
in Figure~\ref{fig:bsllLimits}~\cite{atlascollaboration2021search}.
A set of signal regions (SRs) is defined with lower bounds 
on $m_{\ell\ell}$, $m_{\ell\ell}^\mathrm{min}$, ranging from 400\,GeV to 3200\,(2000)\,GeV for the b-veto (b-tag) category.
Table~\ref{tab:bsll}~\cite{atlascollaboration2021search} lists the
sources of the relative systematic uncertainties for signal regions with $m_{\ell\ell}^\mathrm{min} = 2000~(1500)$\,GeV.

\begin{table}[htb]
\vspace*{-0.7cm}
\caption{
Summary of the relative systematic uncertainties for signal regions with 
$m_{\ell\ell}^\mathrm{min} = 2000~(1500)$\,GeV before the fit is performed for the b-veto (b-tag) categories.~\cite{atlascollaboration2021search}
\label{tab:bsll} }
\begin{center} 
\vspace*{-0.1cm}
\includegraphics[width=\textwidth]{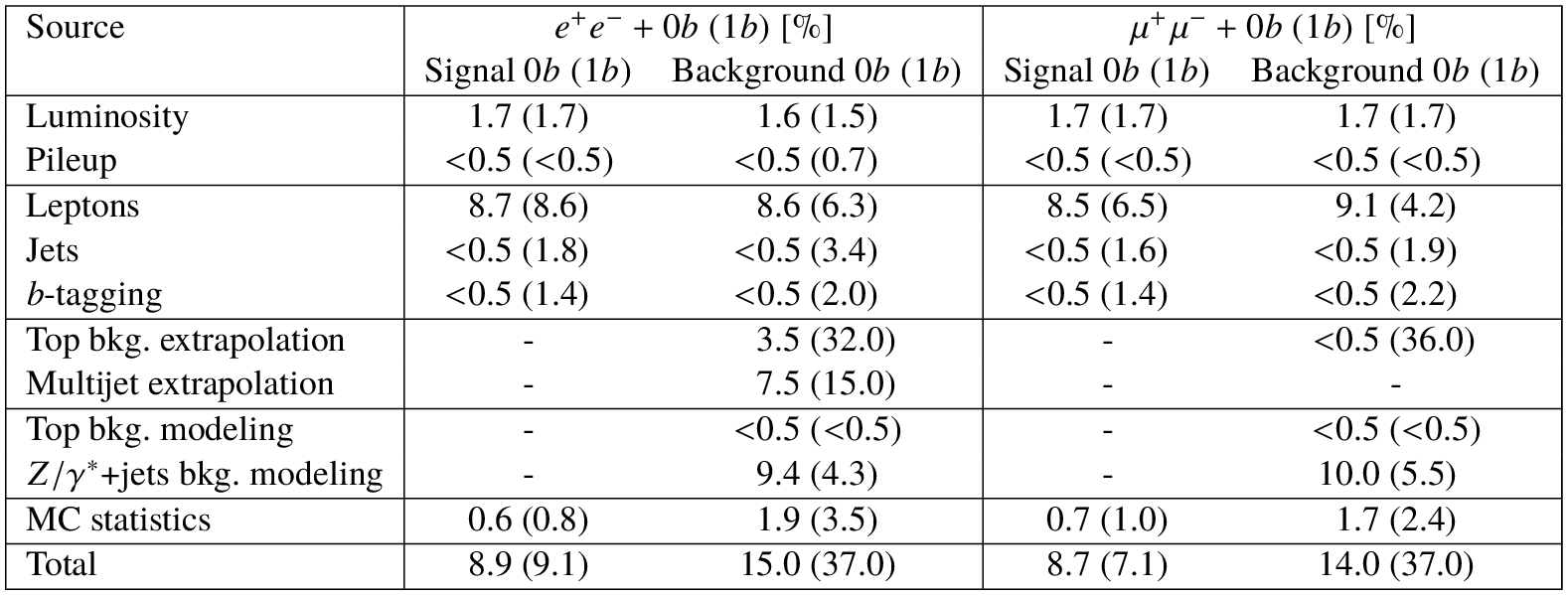}
\end{center}
\vspace*{-0.5cm}
\end{table}

Contact interactions with
$\Lambda/g < 2.0~(2.4)$\,TeV are excluded for ee($\mu\mu$) at 95\%\,CL, still far from the value which is favoured by the B-meson decay anomalies.
Model-independent limits are set as a function of the di-lepton invariant mass for the reinterpretation of the results in terms of other signal scenarios.

\clearpage
\section{Search for 
pair-production of third-generation Leptoquarks in $\mathrm{\mathbf{b\bar b}}$+MET events with taus }
\label{sec:thrid}

\begin{figure}[htb]
\vspace*{-0.2cm}
\begin{center}
\includegraphics[width=0.3\textwidth]{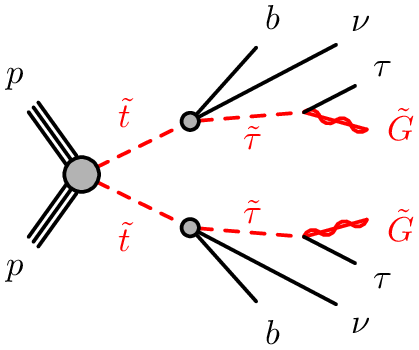}
\includegraphics[width=0.3\textwidth]{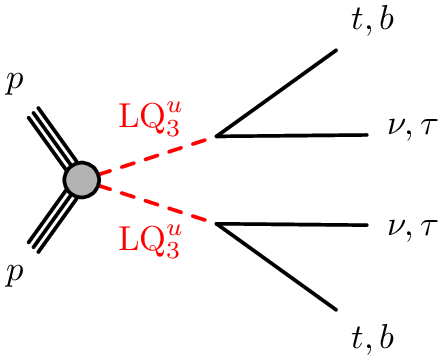}
\includegraphics[width=0.3\textwidth]{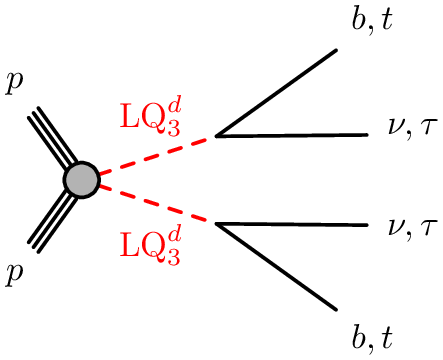}
\end{center}
\vspace*{-0.8cm}
\caption{
Diagrams illustrating the production and decay of particles considered in the simplified models for the Supersymmetric scenario (left) and the scenarios with scalar Leptoquarks of charge $+2/3$e (middle) and $-1/3$e (right).~\cite{ATLAS-CONF-2021-008}
}
\label{fig:bbMETtau}
\end{figure}

Searches for Supersymmetry with scalar top (stop) have sensitivity for Leptoquark pair-production.
In particular the events with b-quarks and missing energy (\bb+MET) together with hadronically decaying tau-leptons are
selected to search for the pair-production of third-generation Leptoquarks~\cite{ATLAS-CONF-2021-008}.
Figure~\ref{fig:bbMETtau}~\cite{ATLAS-CONF-2021-008} shows
the Feynman diagrams for 
scalar-top and Leptoquark production
with a signature of b-quarks, missing
energy and a tau-lepton.
The signatures are 
$\mathrm{
t\nu t\nu,  
b\tau b\tau, 
b\nu b\nu, 
t\tau t\tau.
}$

The search strategy is divided in 
a di-tau and single-tau preselection
as detailed in Table~\ref{tab:bbMETtau}~\cite{ATLAS-CONF-2021-008}.
The further selections are given
in Tables~\ref{tab:bbMETtau2}~\cite{ATLAS-CONF-2021-008}
and~\ref{tab:bbMETtau3}~\cite{ATLAS-CONF-2021-008} for the
di-tau and single-tau preselection,
respectively.

\begin{table}[htb]
\vspace*{-0.2cm}
\caption{
Preselection in the di-tau and single-tau channels.~\cite{ATLAS-CONF-2021-008}
\label{tab:bbMETtau} }
\begin{center} 
\vspace*{-0.1cm}
\includegraphics[width=0.7\textwidth]{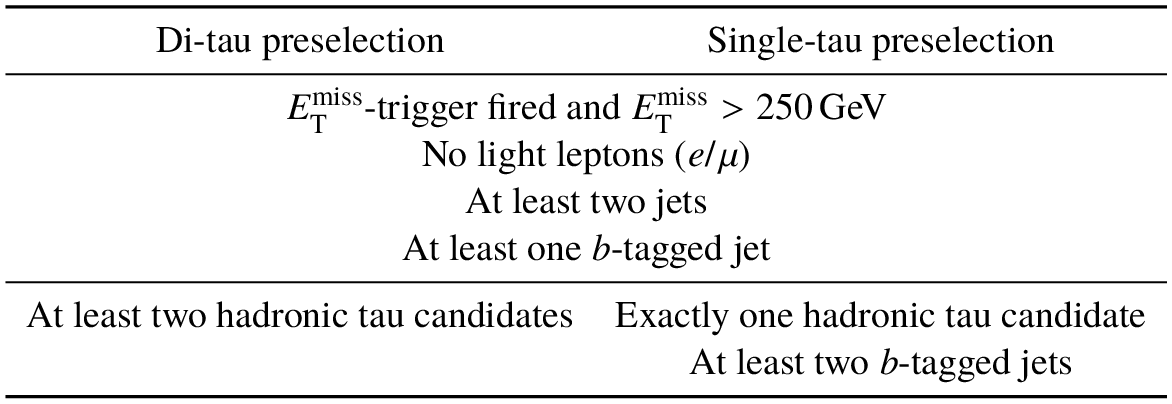}
\end{center}
\vspace*{-0.6cm}
\end{table}

\begin{table}[htb]
\vspace*{-0.2cm}
\caption{
Definitions of the \toptop\ control and validation regions and the signal region in the di-tau channel. A dash signifies that no requirement on the given variable is applied, while brackets indicate an allowed range for the variable. These requirements extend those of the di-tau preselection.~\cite{ATLAS-CONF-2021-008} 
\label{tab:bbMETtau2} }
\begin{center} 
\vspace*{-0.1cm}
\includegraphics[width=\textwidth]{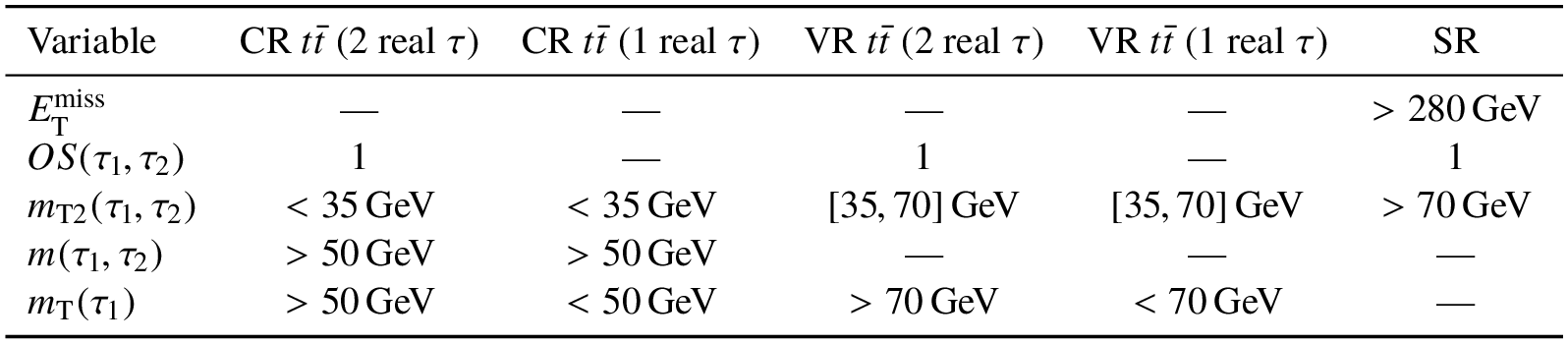}
\end{center}
\end{table}

\begin{table}[htb]
\vspace*{-0.2cm}
\caption{
Definitions of the \bb\
(1 real $\tau$) and single-top control and validation regions and the signal region in the single-tau channel. A dash signifies that no requirement on the given variable is applied, while brackets indicate an allowed range for the variable. In the last column, round brackets enclose the values and ranges used for the multi-bin SR. The binning in $p_T(\tau)$ of the multi-bin SR, abbreviated with "binned", is [50, 100], [100,200], and $>200$\,GeV. These requirements extend those of the single-tau preselection.~\cite{ATLAS-CONF-2021-008}
\label{tab:bbMETtau3} }
\begin{center} 
\vspace*{-0.1cm}
\includegraphics[width=\textwidth]{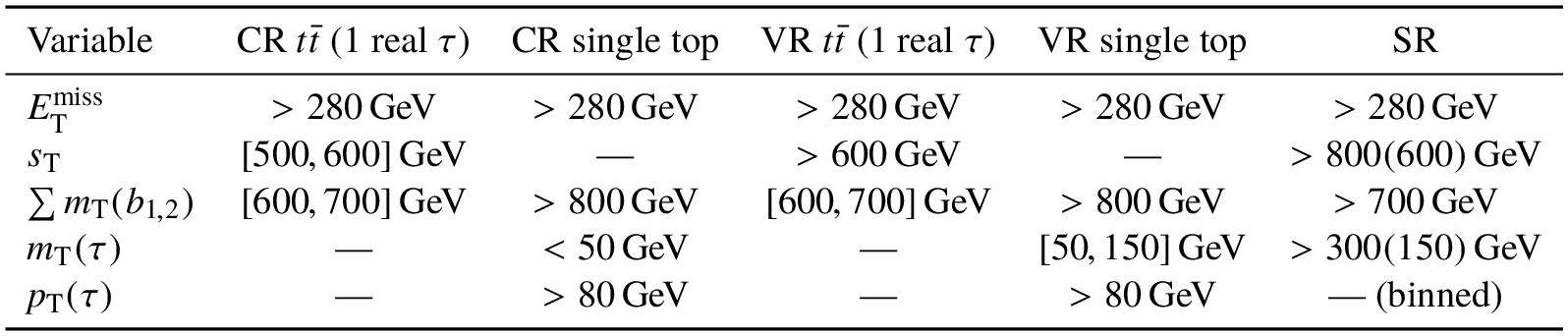}
\end{center}
\vspace*{-0.5cm}
\end{table}

\begin{figure}[htb]
\vspace*{-0.2cm}
\begin{center}
\includegraphics[width=0.49\textwidth]{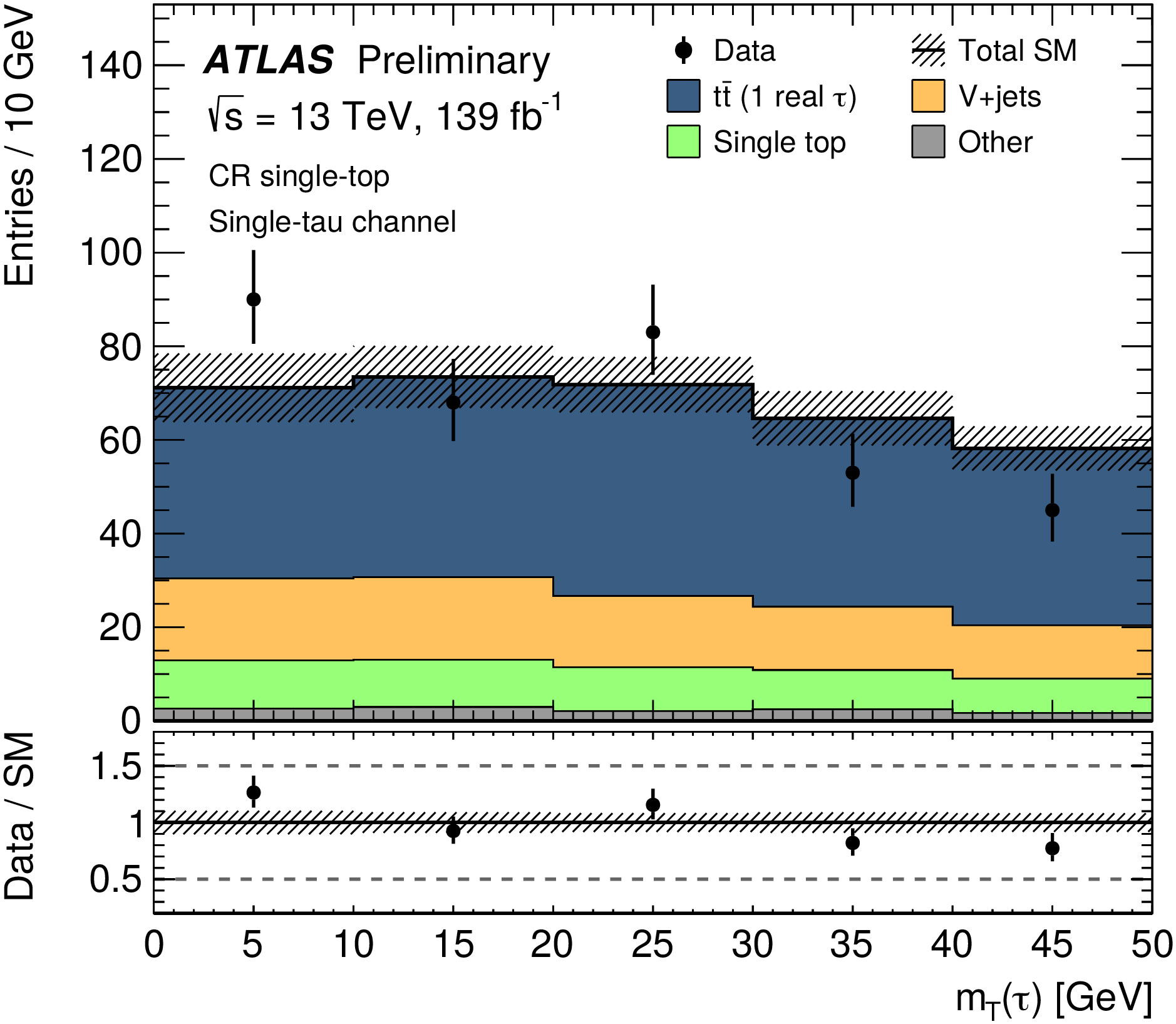}
\includegraphics[width=0.49\textwidth]{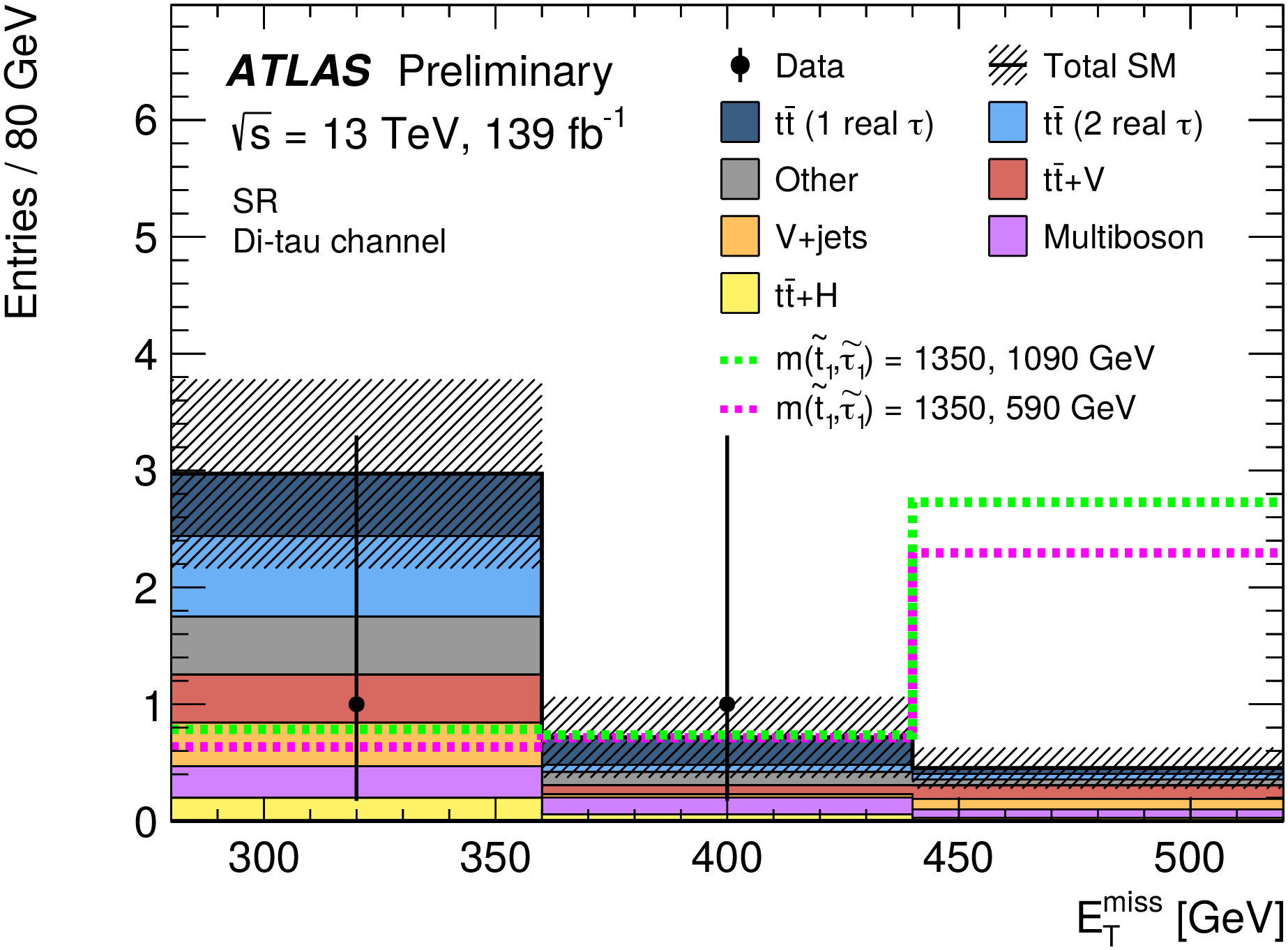}
\end{center}
\vspace*{-0.8cm}
\caption{
Left:
Distributions of $m_T(\tau$)
in the single-top control region of the single-tau channel. The stacked histograms show the various SM background contributions. The hatched band indicates the total statistical and systematic uncertainty of the SM background.  Minor backgrounds are grouped together and denoted as ``Other''. The rightmost bin includes the overflow. 
Right:
Distribution of $E^\mathrm{miss}_T$ 
in the di-tau SR.
The stacked histograms show the various SM background contributions. The hatched band indicates the total statistical and systematic uncertainty of the SM background. The overlaid dotted lines show the additional contributions for signal scenarios close to the expected exclusion contour with the particle type and the mass parameters for the simplified models indicated in the legend.~\cite{ATLAS-CONF-2021-008}
}
\label{fig:bbMETtau2}
\end{figure}

For the single-tau analysis, an example of the data and simulation agreement is shown in Figure~\ref{fig:bbMETtau2}~\cite{ATLAS-CONF-2021-008}
for a control and a signal region.

\begin{figure}[htb]
\vspace*{-0.2cm}
\begin{center}
\includegraphics[width=0.49\textwidth]{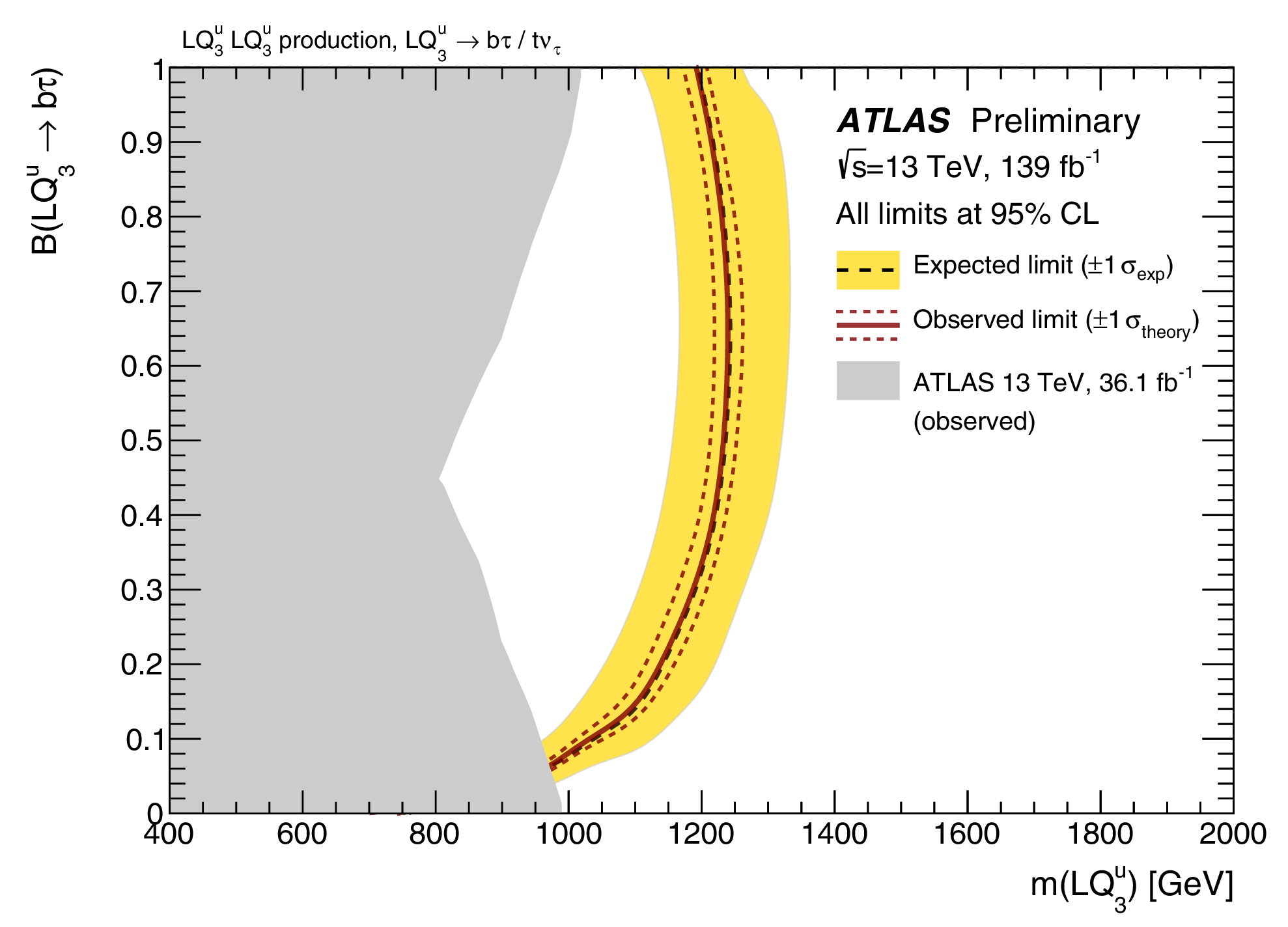}
\includegraphics[width=0.49\textwidth]{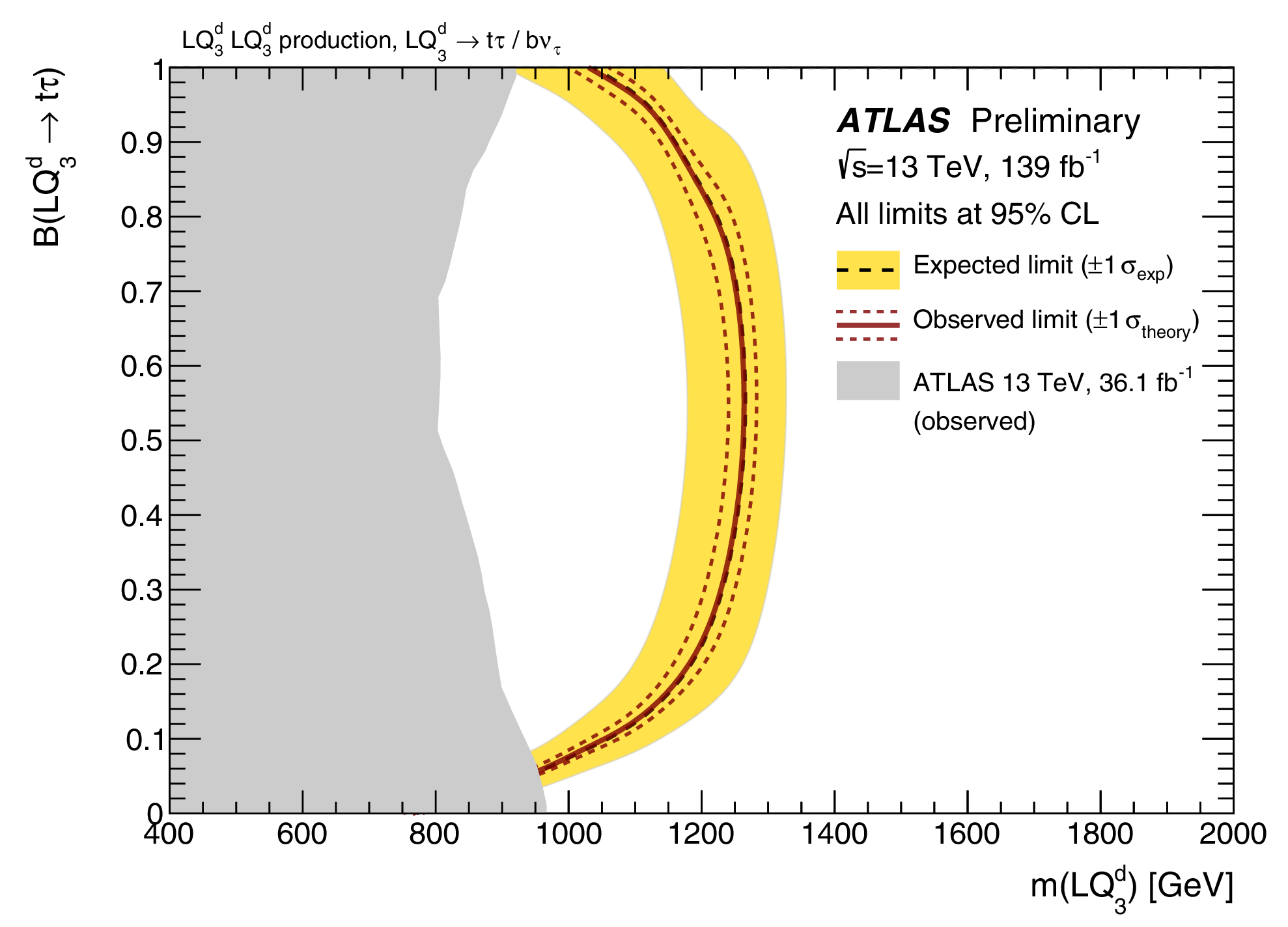}
\end{center}
\vspace*{-0.8cm}
\caption{
Expected and observed exclusion contours at 95\%\,CL for the Leptoquarks signal model, as a function of m(LQ)
 and the branching ratio B($\LQu\rightarrow \mathrm{q}\ell$)
into charged leptons. The left plot shows the exclusion contour for scalar third-generation Leptoquarks \LQu\ with charge $+2/3$e, the right plot the exclusion contour for scalar third-generation Leptoquarks \LQd\ with charge $-1/3$e. The limits are derived from the binned single-tau signal region. Shown in gray for comparison are the observed exclusion-limit contours from the previous ATLAS publication that targets the same Leptoquark models but is based on a subset of the Run-2 data [JHEP 06 (2019) 144]. In this previous publication five different analyses are considered that target not only the final state studied here but also the final states that correspond to a branching fraction of 0 or 1 into a charged lepton and a quark, leading to the concave shapes of the gray exclusion contours.~\cite{ATLAS-CONF-2021-008}
}
\label{fig:bbMETtau3}
\end{figure}

Figure~\ref{fig:bbMETtau3}~\cite{ATLAS-CONF-2021-008} shows expected and observed exclusion contours at 95\%\,CL as a function of $m$(LQ)
and the branching ratio into charged leptons.
For $B(\LQu \rightarrow\mathrm{ b\tau})=0.5$ and 
$B(\LQu \rightarrow \mathrm{t\tau})=0.5$, 
limits for LQs reach 1.25\,TeV.

\section{Search for pair-production of third-generation down-type Leptoquarks in $\mathrm{\mathbf{b\bar b}}$+MET events}
\label{sec:third_down}

\begin{figure}[htb]
\vspace*{-0.2cm}
\begin{center}
\includegraphics[width=0.28\textwidth,]{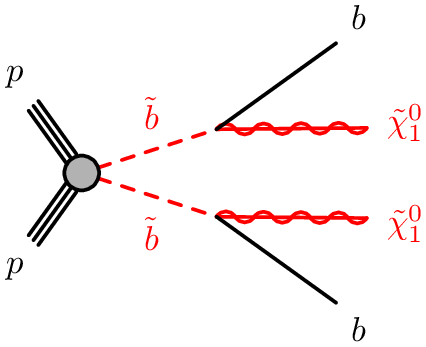}
\includegraphics[width=0.28\textwidth]{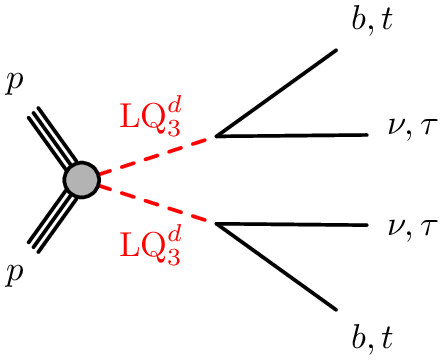}
\includegraphics[width=0.42\textwidth]{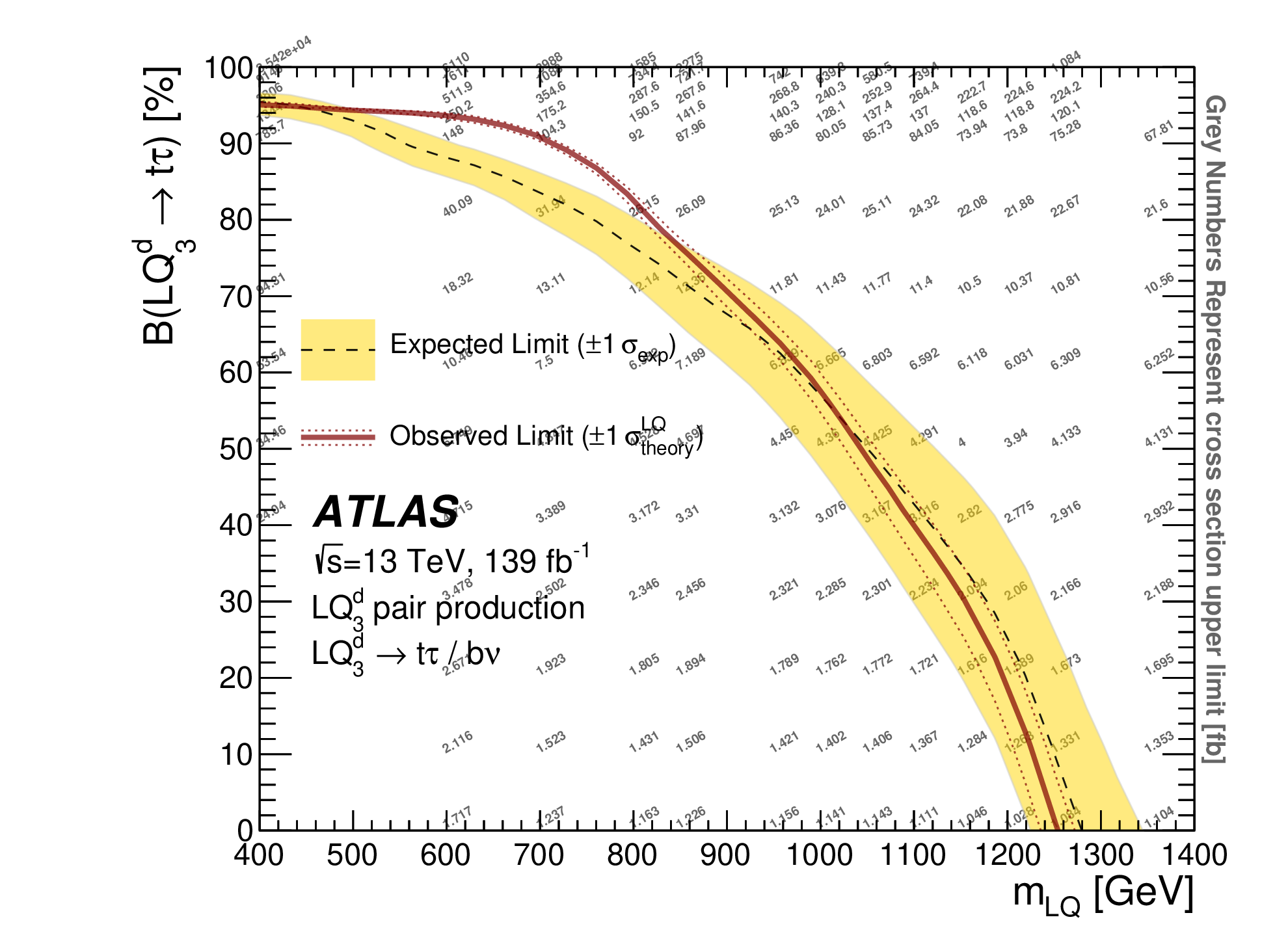}
\end{center}
\vspace*{-0.8cm}
\caption{
Diagrams illustrating the processes targeted by this analysis: (left) bottom squark pair-production,  and (centre) pair-production of scalar third-generation down-type Leptoquarks decaying to bottom quarks and neutrinos or top quarks and $\tau$-leptons. BSM particles are indicated in red, while SM particles are indicated in black.
Right:
Exclusion limits at 95\% CL on B($\mathrm{\LQdthree \rightarrow t\tau}$) as a function of the \LQdthree\ mass for a model of \LQdthree\ pair-production. The dashed line and the shaded band are the expected limit and its $\pm 1\sigma$ uncertainty, respectively. The thick solid line is the observed limit for the central value of the signal cross-section.
The grey numbers show the 95\%\,CL observed exclusion limits on the model cross-section.~\cite{Aad:2750578}
}
\label{fig:third_down}
\vspace*{-0.1cm}
\end{figure}

Searches for \bb+MET Supersymmetric prompt decays have sensitive to pair-production of third-generation LQs.
The corresponding Feynman
graphs are shown in 
Figure~\ref{fig:third_down}~\cite{Aad:2750578}.
Expected and observed mass limits, and cross-section upper limits at 95\%\,CL are also 
give in Figure~\ref{fig:third_down}~\cite{Aad:2750578}.

\section{Search for pair-production of third-generation down-type Leptoquarks in 
$\mathrm{\mathbf{t\bar t}}$+MET all-hadronic events}
\label{sec:ttMEThad}

\begin{figure}[htb]
\vspace*{-0.1cm}
\begin{center}
\includegraphics[width=0.32\textwidth,]{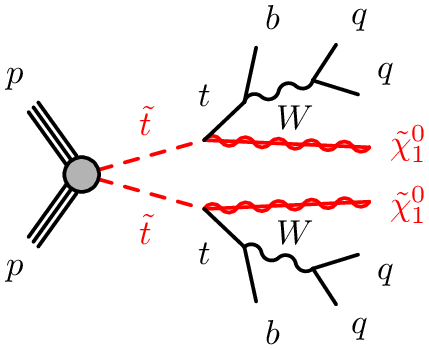}
\includegraphics[width=0.32\textwidth]{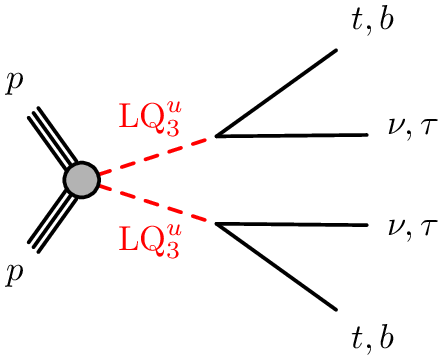}
\includegraphics[width=0.34\textwidth]{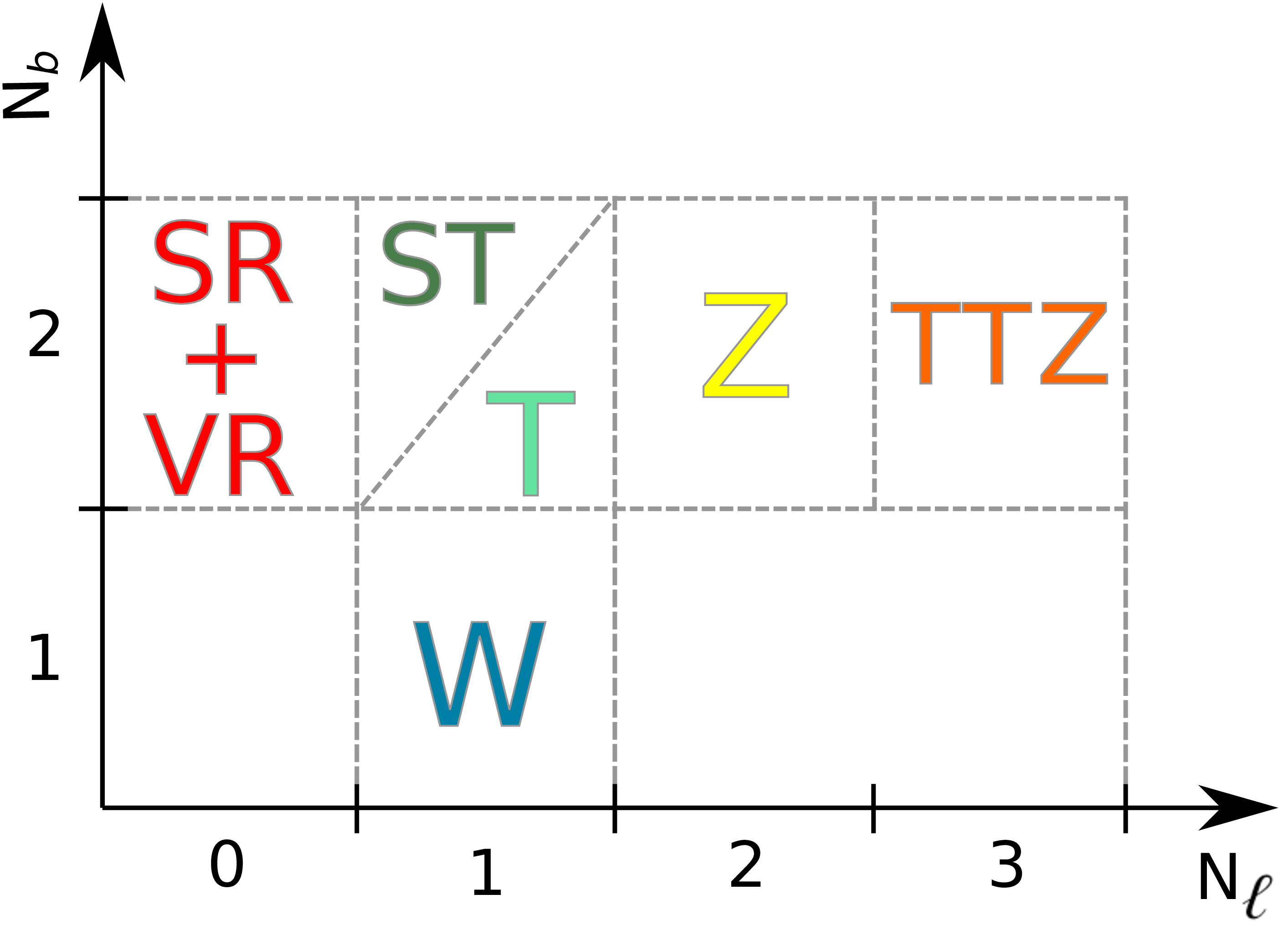}
\end{center}
\vspace*{-0.8cm}
\caption{
Decay topologies of the signal models considered in the analysis: (left) two-body, and (centre) up-type, third-generation scalar Leptoquark pair-production, with both Leptoquarks decaying into a top quark and a neutralino or a bottom quark and a $\tau$-lepton. For simplicity, no distinction is made between particles and antiparticles. Only hadronic W boson decays are shown.
Right:
Summary of background control regions.~\cite{Aad:2716368}
}
\label{fig:third_down_had}
\vspace*{-0.2cm}
\end{figure}

\begin{figure}[htb]
\begin{center}
\includegraphics[width=0.49\textwidth,]{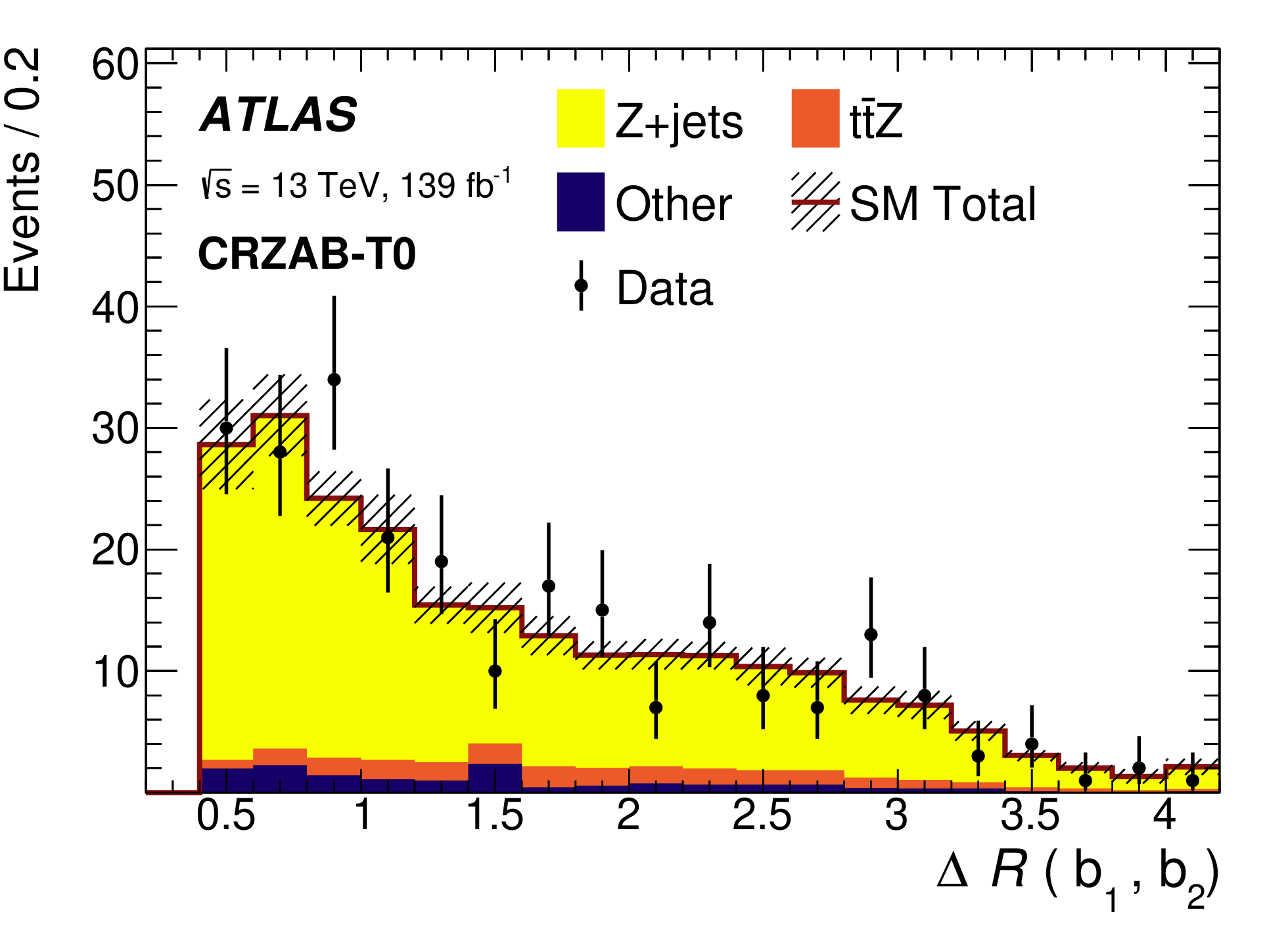}
\includegraphics[width=0.49\textwidth]{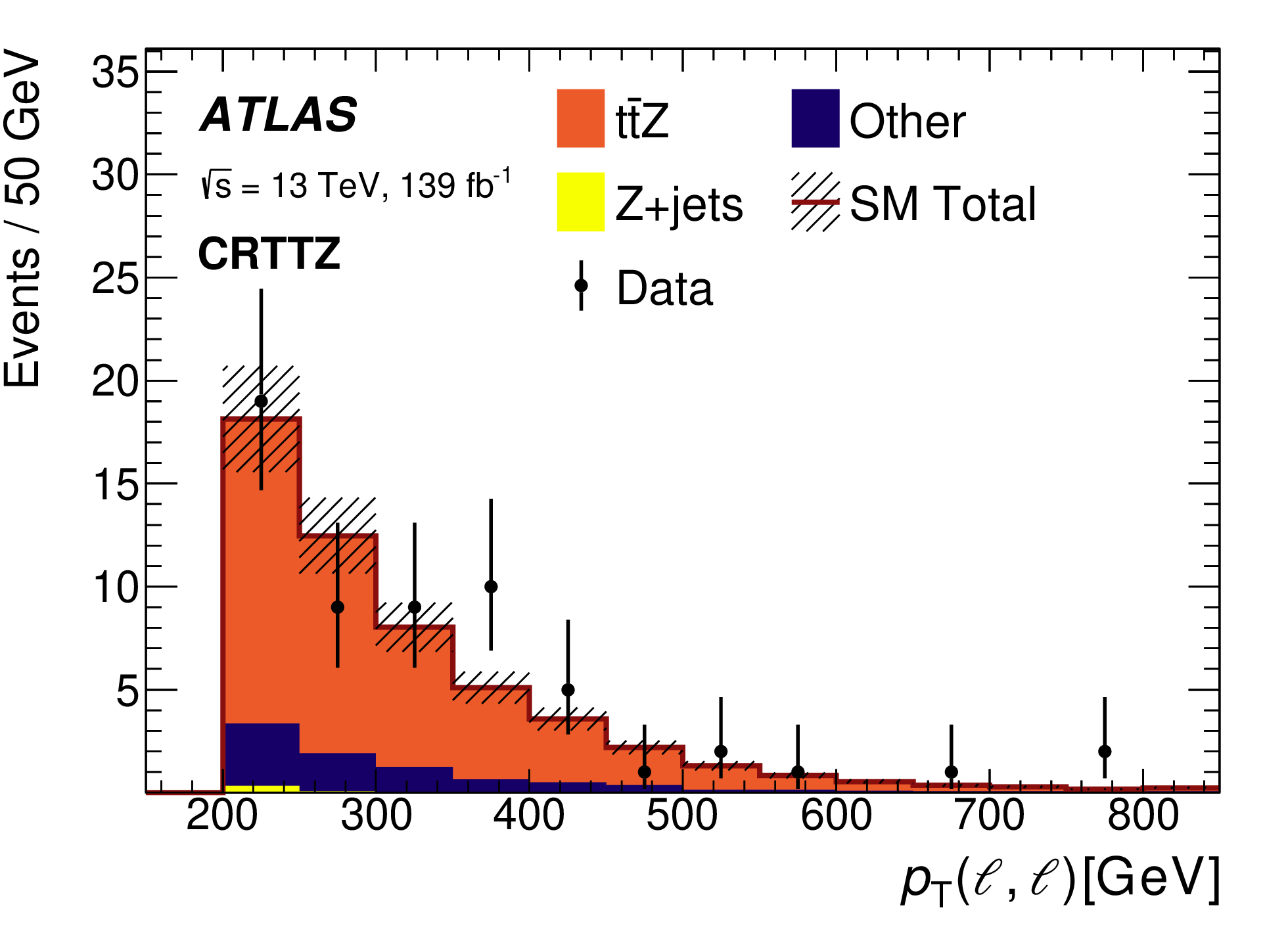}
\end{center}
\vspace*{-0.8cm}
\caption{
Left:
Distributions illustrating the level of agreement between data (points) and the SM expectation (stacked histograms, after simultaneously fitting to all backgrounds) in several Z+jets control regions:
$\Delta R(\mathrm{b1,b2})$ 
for CRZAB-T0. The hatched uncertainty band around the SM expectation includes the combination of MC statistical, theory-related and detector-related systematic uncertainties. The rightmost bin in each plot includes all overflows.
Right:
Distributions illustrating the level of agreement between data (points) and the SM expectation (stacked histograms, after simultaneously fitting to all backgrounds) in the
\toptop+Z
control region:  pT($\ell,\ell$) 
for CRTTZ.~\cite{Aad:2716368}
}
\label{fig:third_down_had2}
\end{figure}

\begin{figure}[htb]
\vspace*{-0.1cm}
\begin{center}
\includegraphics[width=0.49\textwidth,]{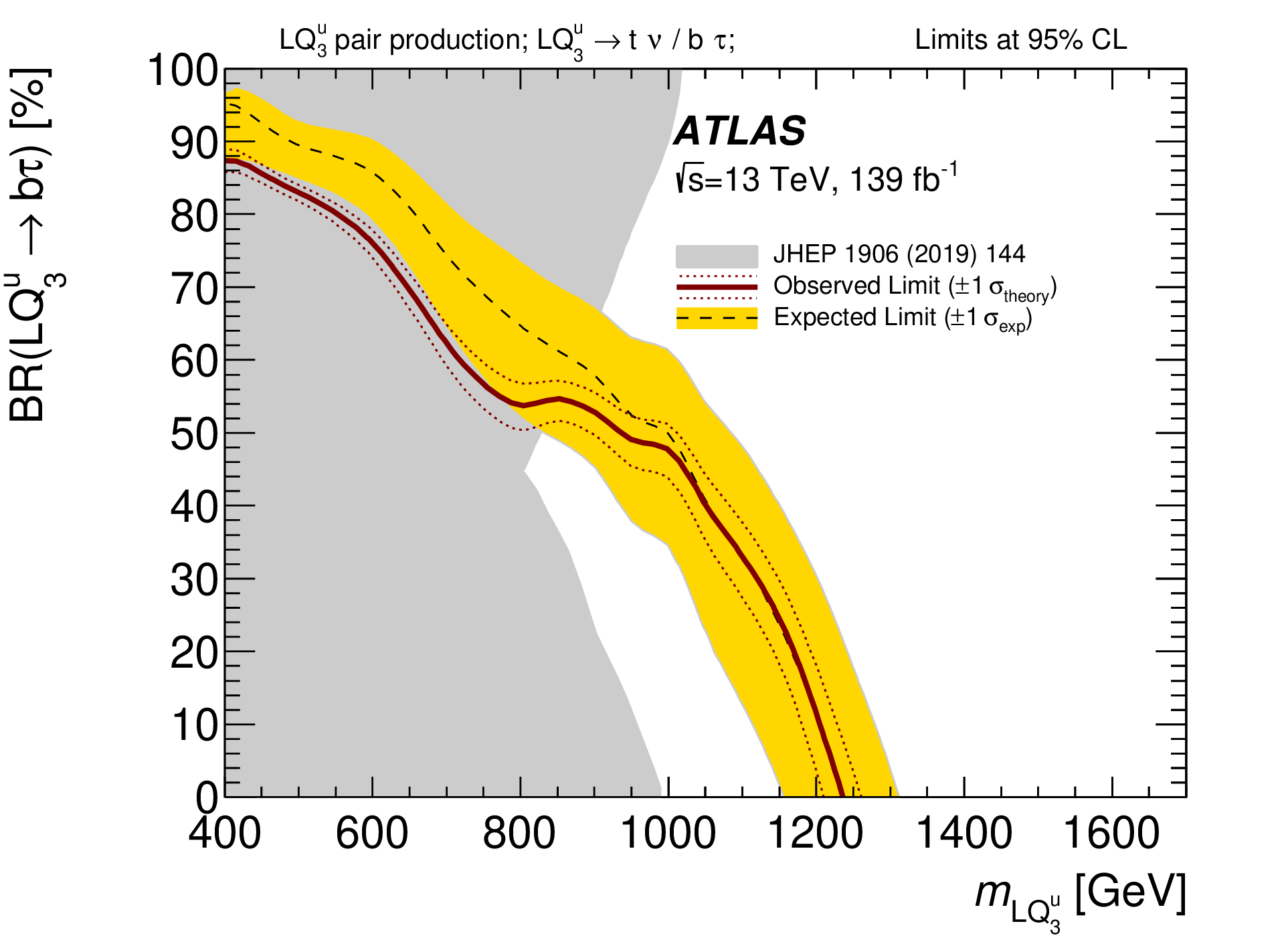}
\includegraphics[width=0.49\textwidth]{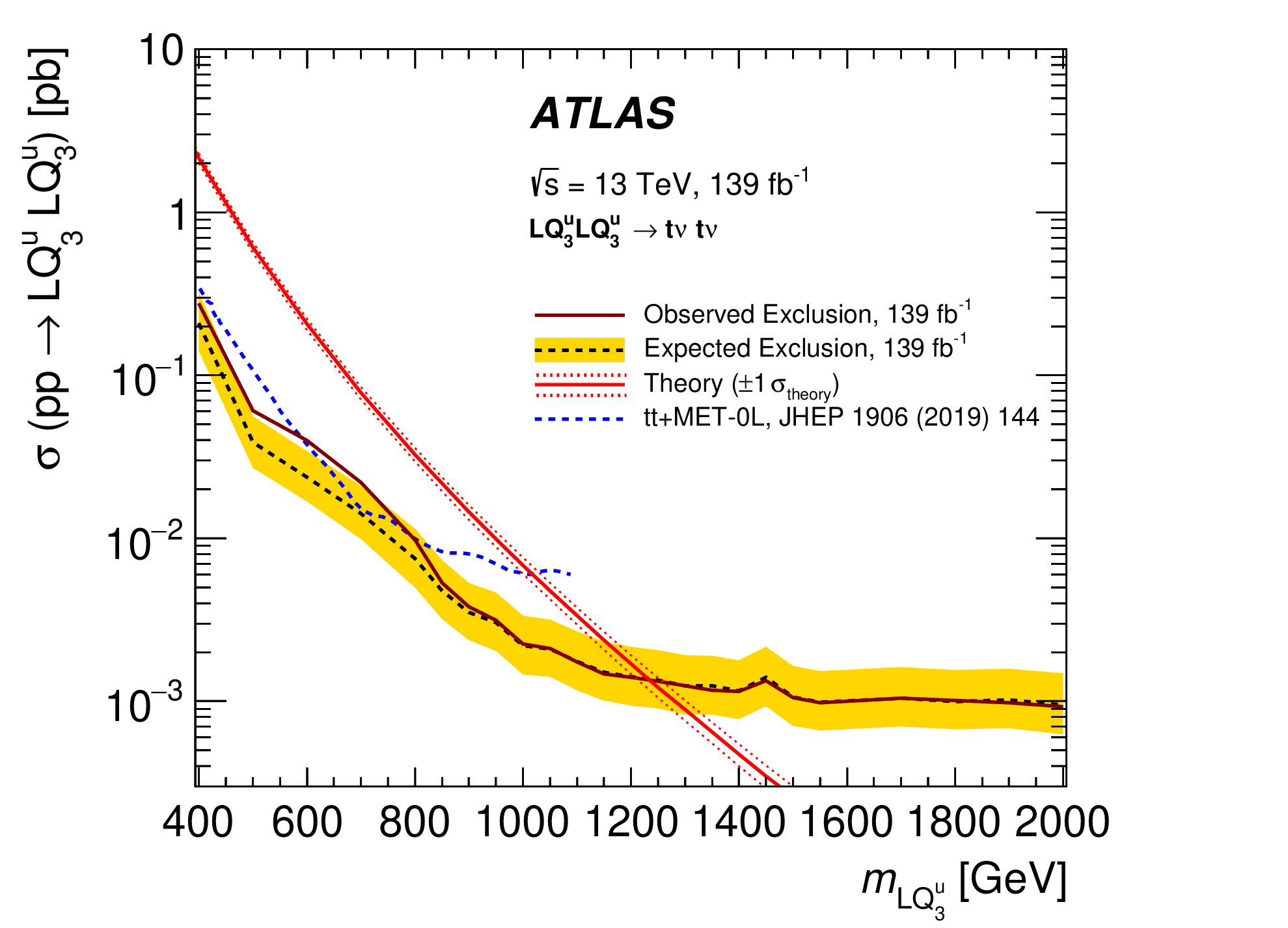}
\end{center}
\vspace*{-0.8cm}
\caption{
Observed (red solid line) and expected (black dashed line) limits on up-type, third-generation Leptoquarks. Left: Limits as a function of the branching ratio of Leptoquarks decaying into 
$\mathrm{b\tau}$
(with the only other decay allowed being into
$\mathrm{t\nu}$) as a function of
the Leptoquark mass. 
Right: Limits on the production cross-section at 95\% CL as a function of Leptoquark mass assuming that all Leptoquarks decay into 
$\mathrm{t\nu}$. Uncertainty bands corresponding to the
$\pm 1\sigma$ 
variation of the expected limit (yellow band) and the sensitivity of the observed limit are also indicated. Observed limits from previous searches with the ATLAS detector at $\sqrt s=13$\,TeV 
are overlaid (left) in grey and (right) as a blue dashed line.~\cite{Aad:2716368}
}
\label{fig:third_down_had3}
\end{figure}

Searches for \toptop+MET Supersymmetric prompt decays have sensitivity to pair-production of third-generation Leptoquarks.
The corresponding Feynman
graphs are shown in 
Figure~\ref{fig:third_down_had}~\cite{Aad:2716368}.
The control regions for
Z+jets (Z), \toptop Z (TTZ), \toptop~(T), W+jets (W), and single-top (ST) backgrounds
are ordered according to the 
numbers of leptons and b-jets,
as also shown in Figure~\ref{fig:third_down_had}~\cite{Aad:2716368}.
Examples of the control regions 
for Z+jets and \toptop Z are shown in
Figure~\ref{fig:third_down_had2}~\cite{Aad:2716368}.
Excluded \LQuthree\ (masses, branching ratios) and cross-section limits for 
\LQuthree\ pair-production
are given in Figure~\ref{fig:third_down_had3}~\cite{Aad:2716368}.

\clearpage
\section{Summary of results}
\label{sec:summary}

\begin{figure}[htb]
\vspace*{-0.2cm}
\begin{center}
\includegraphics[width=0.49\textwidth]{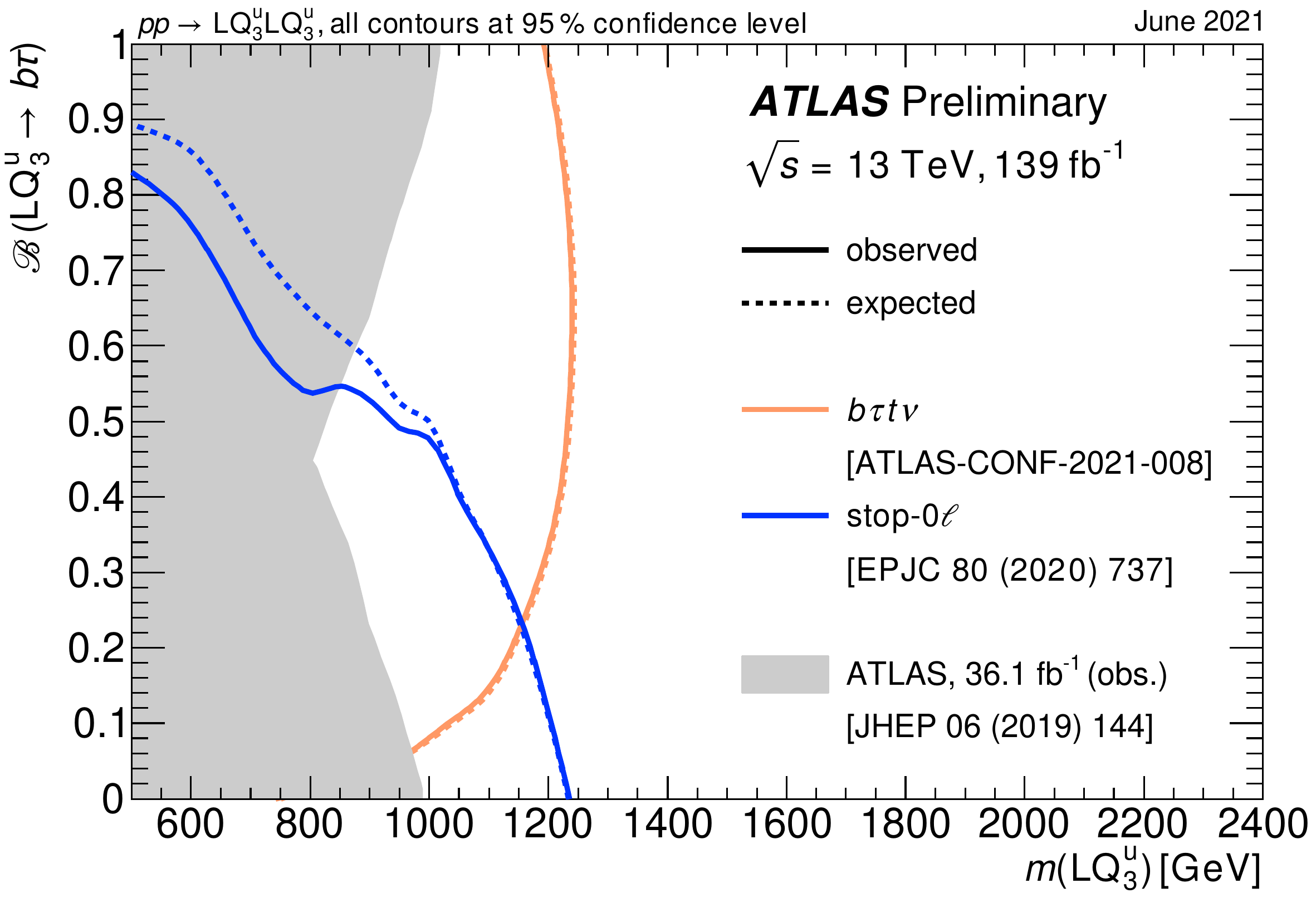} \hfill
\includegraphics[width=0.49\textwidth]{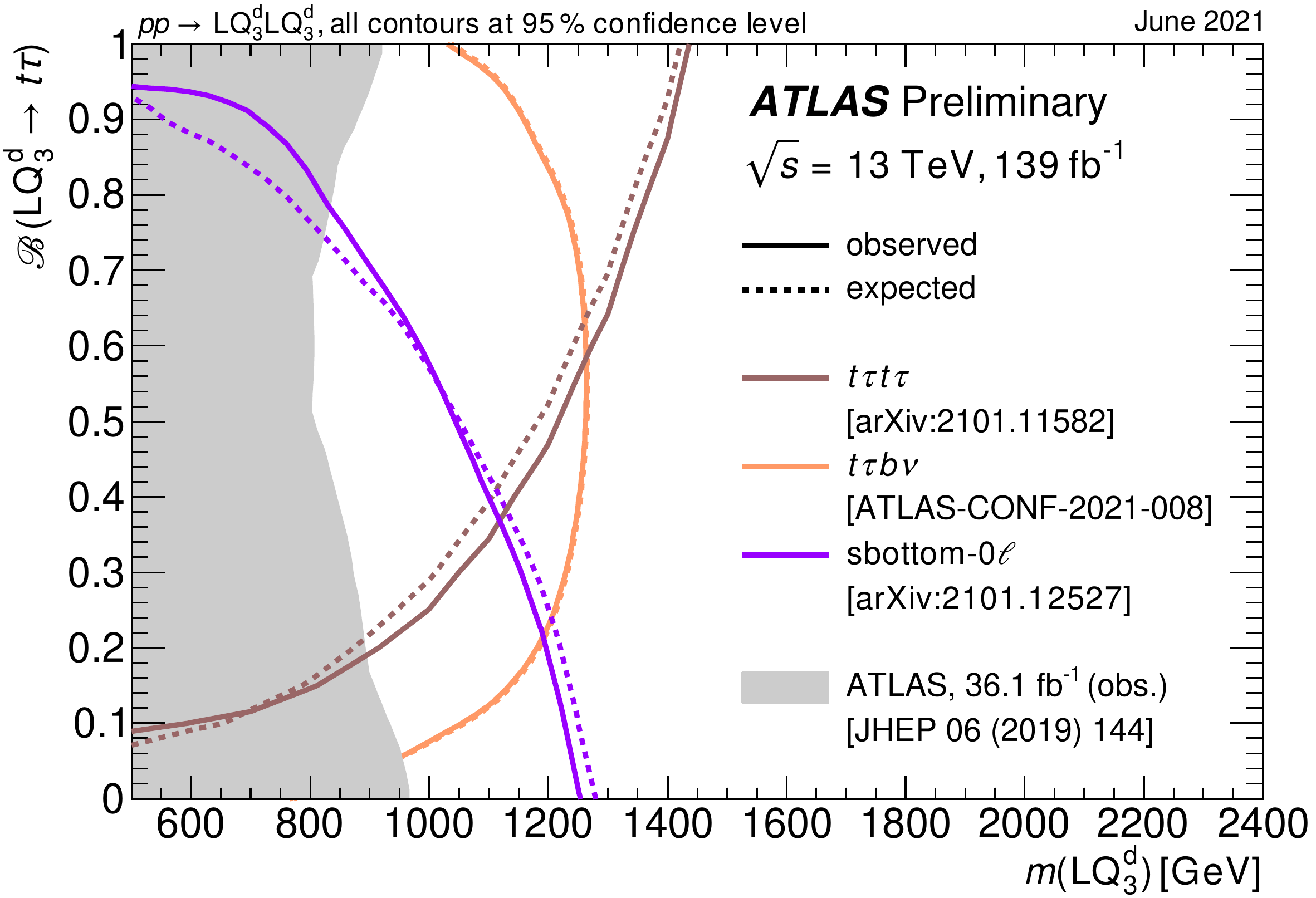}
\end{center}
\vspace*{-0.8cm}
\caption{
Left: 
Expected and observed exclusion contours at the 95\%\,CL for pair-produced scalar third-generation up-type Leptoquarks with decays 
$\LQuthree\rightarrow \mathrm{t\tau / \mathrm{b\nu}}$, as a function of the Leptoquark mass and the branching fraction
B($\LQuthree \rightarrow \mathrm{b\tau}$) into a charged lepton and a quark. In addition to the dedicated search for Leptoquarks, the plot includes a reinterpretation of the search for pair-production of Supersymmetric top squarks with no leptons (stop-0$\ell$) in the final state.
Right:
Expected and observed exclusion contours at the 95\%\,CL for pair-produced scalar third-generation down-type Leptoquarks with decays $\LQdthree\rightarrow \mathrm{b\nu / t\tau}$, as a function of the Leptoquark mass and the branching fraction B($\LQdthree \rightarrow \mathrm{t\tau}$) into a charged lepton and a quark. In addition to the dedicated searches for Leptoquarks, the plot includes a reinterpretation of the search for pair-production of Supersymmetric bottom squarks with no leptons (sbottom-0$\ell$) in the final state.~\cite{ATL-PHYS-PUB-2021-017}
}
\label{fig:summary1}
\vspace*{-0.1cm}
\end{figure}

\begin{figure}[htb]
\vspace*{-0.2cm}
\begin{center}
\includegraphics[width=0.49\textwidth]{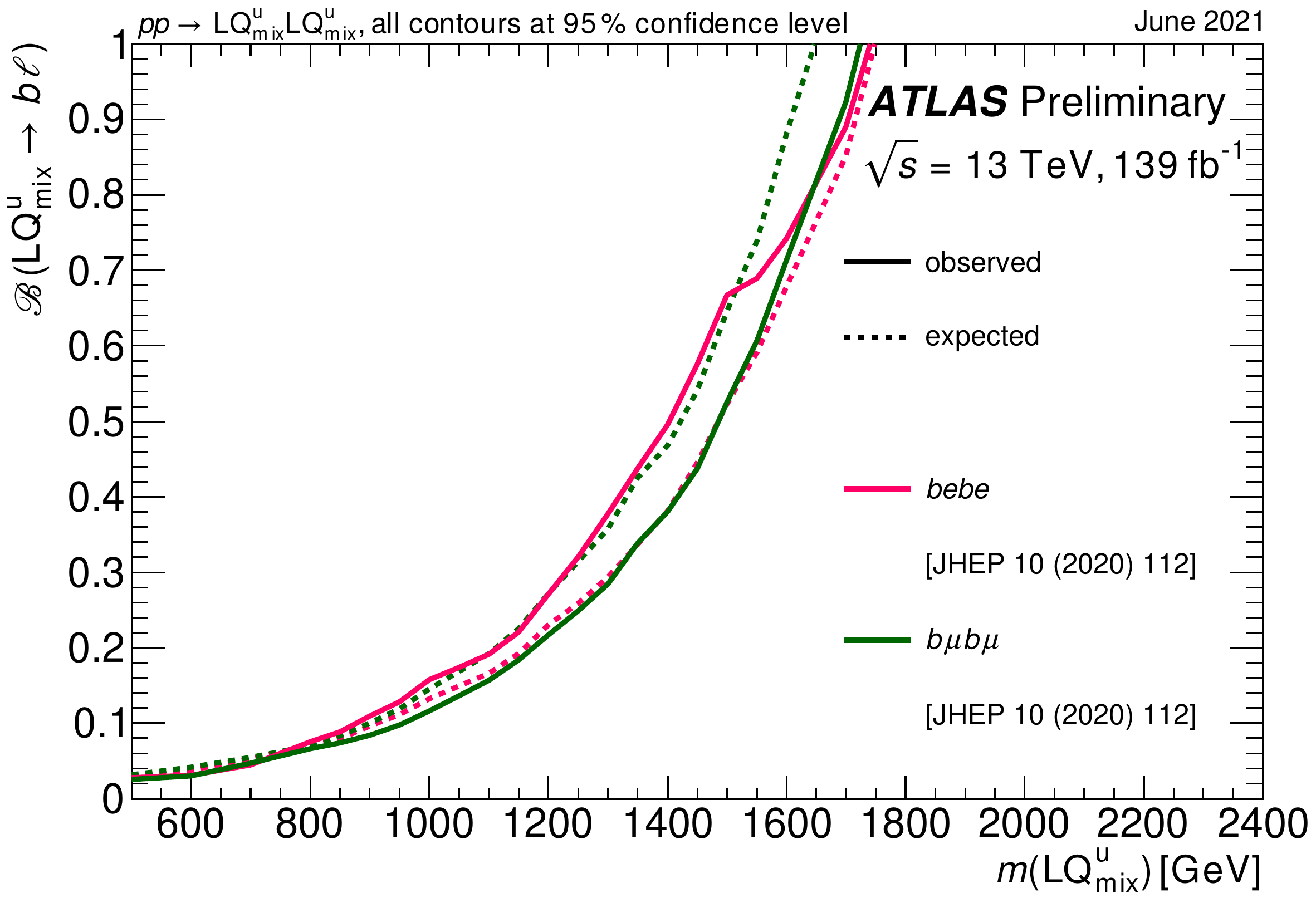} \hfill
\includegraphics[width=0.49\textwidth]{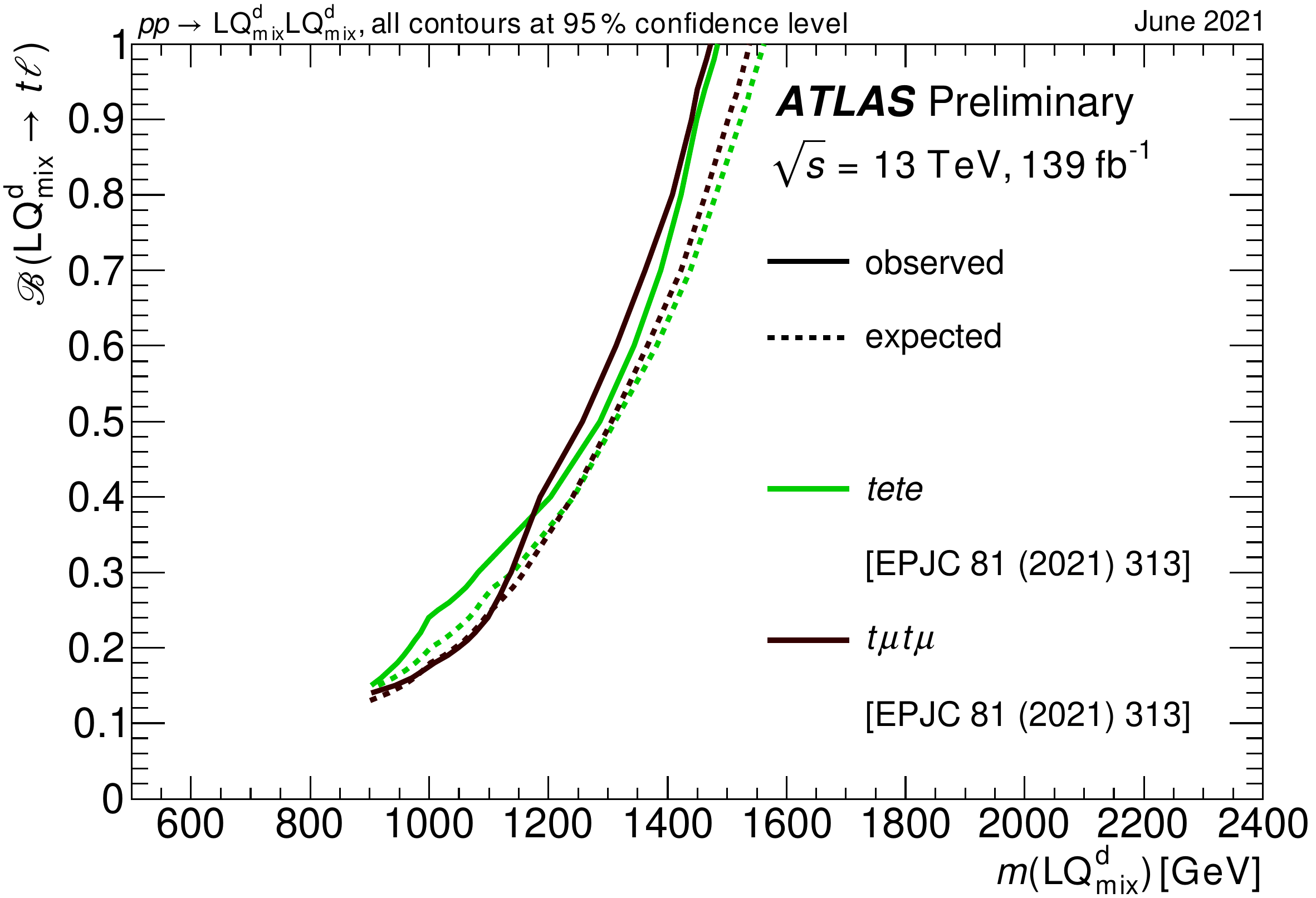}
\end{center}
\vspace*{-0.8cm}
\caption{
Left:
Expected and observed exclusion contours at the 95\%\,CL for pair-produced scalar up-type Leptoquarks with mixed decays to third-generation quarks and leptons from the first or second generation 
$\LQumix\rightarrow \mathrm{t\tau / \mathrm{b\ell}}$, as a function of the Leptoquark mass and the branching fraction 
B($\LQumix \rightarrow \mathrm{b\ell}$) into a charged lepton and a quark. 
Right:
Expected and observed exclusion contours at the 95\%\,CL for pair-produced scalar down-type Leptoquarks with mixed decays to third-generation quarks and leptons from the first or second generation 
$\LQdmix\rightarrow \mathrm{b\nu / t\ell}$, as a function of the Leptoquark mass and the branching fraction 
B($\LQdmix \rightarrow \mathrm{t\ell}$) into a charged lepton and a quark. The search shown here selects final states with boosted hadronically decaying top quarks and thus probes high Leptoquark masses above 900 GeV.~\cite{ATL-PHYS-PUB-2021-017}
}
\label{fig:summary2}
\vspace*{-0.1cm}
\end{figure}

Up-type and down-type
third-generation model 
Leptoquark
(\LQuthree\ and \LQdthree) 
results are summarized in
Figure~\ref{fig:summary1}~\cite{ATL-PHYS-PUB-2021-017}.
The corresponding mixed-generation
results are given in 
Figure~\ref{fig:summary2}~\cite{ATL-PHYS-PUB-2021-017}.

\clearpage
An overview of the ATLAS results from searches for Leptoquarks
are given in Table~\ref{tab:summary}.

\begin{table}[htb]
\caption{
Summary of Leptoquark searches with ATLAS data. The lower mass limits are given in TeV at 95\%\,CL.
\label{tab:summary} }
\begin{center} 
\vspace*{-0.1cm}
\includegraphics[width=\textwidth]{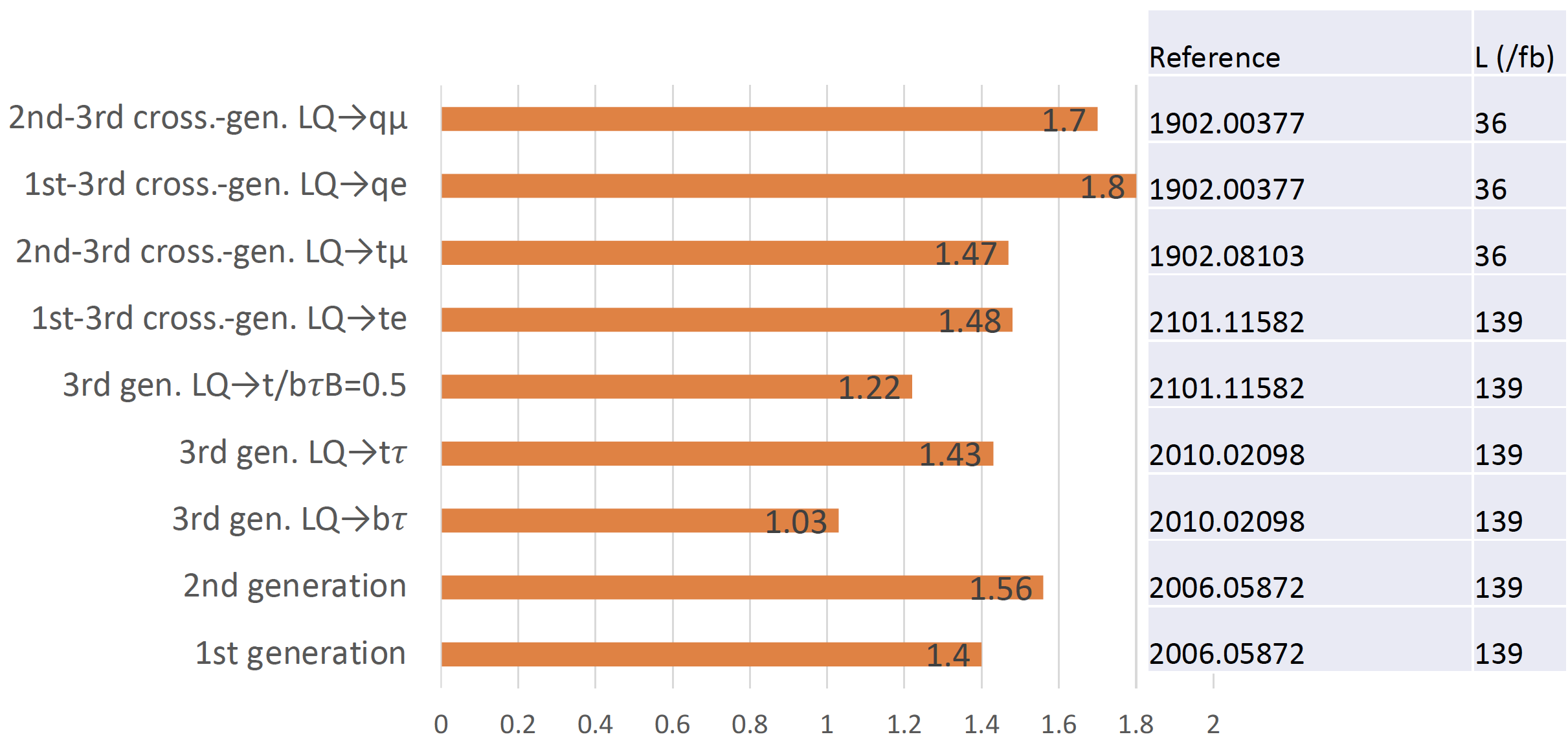}
\end{center}
\end{table}

\section{Leptoquark and Higgs boson searches}

\begin{figure}[htb]
\vspace*{-0.3cm}
\begin{center}
\includegraphics[width=0.4\textwidth]{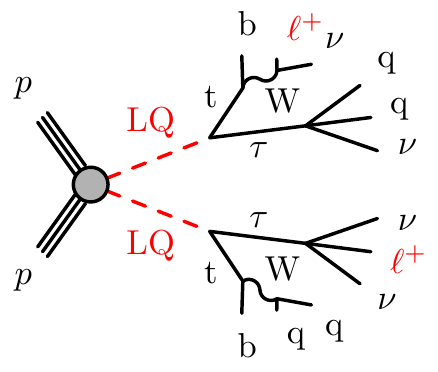}\hfill
\includegraphics[width=0.4\textwidth]{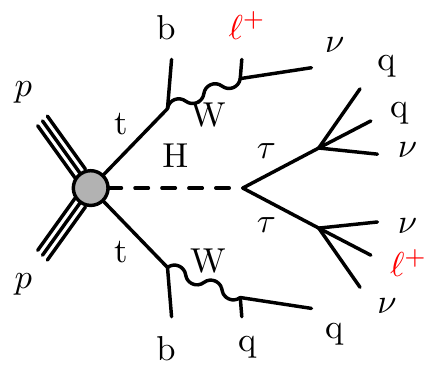}
\end{center}
\vspace*{-0.8cm}
\caption{
Illustration that the same final state 
for Higgs boson production and Leptoquark production is expected.
}
\label{fig:summary3}
\vspace*{-0.8cm}
\end{figure}

As a novel idea it is noted that 
there are also possibilities to 
reinterpret results from Higgs boson searches as limits on Leptoquarks.
As an example, the final state with two light leptons and one hadronically decaying tau is given.
This final state was studied in
$\mathrm{\toptop H(H\rightarrow \tau\tau)}$~\cite{ATLAS-CONF-2019-045}
production, and it has the
same final state as expected in Leptoquark production (Figure~\ref{fig:summary3}).

\clearpage
\section{Conclusions}
\label{sec:conclusions}
There is growing interest in Leptoquarks as a possible explanation of the recently observed B-anomaly (hints for lepton flavour universality violation).
Contact interaction limits $\Lambda /g > 2.0 (2.4)$\,TeV were set at 95\% CL for ee($\mu\mu$) final states. 
These are not sensitive yet to probe the suggested range by B-anomaly ($\sim 30$\,TeV).
Model-independent limits are 
set as a function of the di-lepton invariant mass.
Direct searches for Leptoquarks and re-interpretations of searches for Supersymmetry exclude phase-space of 1st, 2nd and 3rd generation Leptoquarks. 
The current focus is on the 3rd generation, including cross-generational decays.
The potential for reinterpretations of Higgs boson results is highlighted.
Search results are largely statistically limited, thus more sensitivity is expected with new data.
There is a large potential in flavour physics for collaborations of phenomenologists and experimentalists.

\section*{Acknowledgments}
The project is supported by the 
Ministry of Education, Youth and Sports of the Czech Republic under project number LTT\,17018.

Copyright 2021 CERN for the benefit of the ATLAS Collaboration. CC-BY-4.0 license.

\section*{References}

\bibliographystyle{unsrt}

\bibliography{pheno21}

\end{document}